\documentclass[showpacs,showkeys,nofootinbib]{revtex4}%
\usepackage{graphicx}
\usepackage{epsfig}
\usepackage{bm}
\usepackage{amsmath}
\usepackage{amsfonts}
\usepackage{amssymb}
\usepackage{subfigure}%
\makeatletter
\newcommand\erfc{\mathop{\operator@font erfc}\nolimits}
\def\slashchar#1{\setbox0=\hbox{$#1$}
\dimen0=\wd0 \setbox1=\hbox{/} \dimen1=\wd1
\ifdim\dimen0>\dimen1 \rlap{\hbox to \dimen0{\hfil/\hfil}} #1
\else  \rlap{\hbox to \dimen1{\hfil$#1$\hfil}} / \fi}

\makeatother
\begin{document}
\title{Photon distribution amplitudes and light-cone wave functions in chiral quark
models\thanks{Research supported by the Polish State Committee for Scientific
Research, grant 2P03B~02828}}
\author{Alexander E. Dorokhov}
\email{dorokhov@theor.jinr.ru}
\affiliation{Joint Institute for Nuclear Research, Bogoliubov Laboratory of Theoretical
Physics, 114980, Moscow region, Dubna, Russia}
\author{Wojciech Broniowski}
\email{Wojciech.Broniowski@ifj.edu.pl}
\affiliation{Institute of Physics, \'Swi\c{e}tokrzyska Academy, PL-25406~Kielce, Poland}
\affiliation{H. Niewodnicza\'nski Institute of Muclear Physics, PL-31342~Krak\'ow, Poland}
\author{Enrique Ruiz Arriola}
\email{earriola@ugr.es}
\affiliation{Departamento de F\'{\i}sica At\'omica, Molecular y Nuclear, Universidad de
Granada, E-18071 Granada, Spain}
\date{July 14, 2006}

\begin{abstract}
The leading- and higher-twist distribution amplitudes and light-cone wave
functions of real and virtual photons are analyzed in chiral quark models. The
calculations are performed in the nonlocal quark model based on the instanton
picture of QCD vacuum, as well as in the spectral quark model and the
Nambu--Jona-Lasinio model with the Pauli-Villars regulator, which both treat
interaction of quarks with external fields locally. We find that in
all considered models
the leading-twist distribution amplitudes of the real photon
defined at the quark-model momentum scale are constant or remarkably close to the constant
in the $x$ variable, thus are far from the asymptotic
limit form. The QCD evolution to higher momentum scales is necessary and we
carry it out at the leading order of the perturbative theory for the
leading-twist amplitudes. We provide estimates for the magnetic susceptibility
of the quark condensate $\chi_{\mathrm{m}}$ and the coupling $f_{3\gamma}$,
which in the nonlocal model turn out to be close to the estimates from QCD sum rules.
We find the higher-twist
distribution amplitudes at the quark model scale and compare them to the Wandzura-Wilczek
estimates. In addition, in the
spectral model we evaluate the distribution amplitudes and light-cone
wave functions of the $\rho$-meson.

\end{abstract}


\maketitle

\section{ Introduction}

Investigations of hard exclusive processes are essential for our understanding
of the internal quark-gluon dynamics of hadrons. Theoretically, such studies
are based on the assumption of factorization of dynamics at long and short
distances. The short-distance physics is well elaborated by perturbative
methods of QCD and depends on particular hard subprocesses. The long-distance
dynamics is essentially nonperturbative and within the factorization formalism
becomes parametrized in terms of hadronic \emph{distribution amplitudes} (DAs)
or their transverse-momentum unintegrated generalizations, the
\emph{light-cone wave functions} (LCWF)\footnote{For a recent review see,
\emph{e.g.}, \cite{Belitsky:2005qn}.}. These nonperturbative quantities are
universal and are defined as vacuum-to-hadron matrix elements of particular
nonlocal light-cone quark or quark-gluon operators. The evolution of DAs at
sufficiently large virtuality $q^{2}$ is controlled by the renormalization
scale dependence of the quark bilinear operators within the QCD perturbation
theory. For leading-order DAs this dependence is governed by simple QCD
evolution equations of the Efremov-Radyushkin-Brodsky-Lepage (ER-BL) type
\cite{Radyushkin:1977gp,Efremov:1978rn,Efremov:1979qk,Lepage:1979zb,Lepage:1980fj}%
. When the normalization scale goes to infinity the DAs reach an ultraviolet
fixed point and are uniquely determined by perturbative QCD. However, the
derivation of the DAs themselves at an initial scale $\mu_{0}^{2}$ from first
principles is a nonperturbative problem and remains a serious challenge.
Moreover, at experimentally achievable energies it is likely that the
normalization scale is low and the nonperturbative effects are essential. It
is expected that at such low scales the shape of DAs can differ essentially
from their asymptotic forms and at present can not be strictly predicted from
first principles. Nevertheless, there are several well-established methods
to get information on genuinely nonperturbative quantities such as hadronic
distribution amplitudes at low normalization scale: the QCD sum rules method
\cite{Chernyak:1981zz,Mikhailov:1986be,Braun:1988qv,Bakulev:1994su,Agaev:2004dc}%
, the relativistic quark models \cite{Bagdasaryan:1984kz}, the instanton
liquid model
\cite{Esaibegian:1989uj,Dorokhov:1991nj,Dorokhov:2000gu,Petrov:1998kg,Dorokhov:2002iu}%
, the effective chiral quark models
\cite{Praszalowicz:2001wy,RuizArriola:2002bp}, or transverse lattice QCD
\cite{Dalley:2002nj}. First and second moments of pion DAs have also been
evaluated on Euclidean lattices
\cite{DelDebbio:2002mq,DelDebbio:2005bg,Gockeler:2005jz}.

A special class of processes involves either virtual or real photons which
provide ideal tools for probing the hadronic structure in deep-inelastic and
hard exclusive scattering experiments. The reason is that the photon is not an
eigenstate of QCD but a superposition of the $U_{1}$ gauge boson and
quark-gluon configurations which are suppressed by the electromagnetic
coupling. While a photon is normally considered a structureless particle, it
can fluctuate into a charged lepton- and quark-pair states, which can be
revealed through interactions with a highly virtual photon. Thus, the photon
DAs have both electromagnetic and hadronic components. The electromagnetic
component can be calculated within QED, while the hadronic part must be
analyzed with non-perturbative methods of strong interactions.

Despite the wealth of theoretical studies not much experimental information is
available to constrain various theoretical predictions
(for a recent review see \cite{Ashery:2006zw}). The hadronic DAs and
LCWFs can be measured from hard exclusive processes at some high momentum
scale. For example the pion and photon DAs may be extracted from hard dijet
production by incident pions \cite{Nikolaev:2000sh,Braun:2001ih} and real
photons \cite{Brodsky:1994kf,Braun:2002en}, correspondingly. The measurements
of the photon DA and LCWF are based on the method of diffractive dissociation
applied already to the study of the pion distributions. A differential
measurement of the pion DA and LCWF was performed by the Fermilab E791
collaboration by studying the diffractive dissociation of high momentum pions
into two jets \cite{Aitala:2000hc}. The recent measurements at HERA provided
the first evidence that diffractive dissociation of particles can be reliably
used to measure the photon DAs. Hence, the electromagnetic component of the
photon LCWF may be extracted from the study of the exclusive $ep\rightarrow
ep\mu^{+}\mu^{-}$ photoproduction process \cite{Ukleja:2004yt} and the
hadronic component from the exclusive processes of diffractive photo- or
electro-production of two pions \cite{Coppola:2004br}. This may be considered
as a special case of the photon dissociation to dijets when each jet consists
of one pion.

The aim of the present work is to study the photon DAs and LCWFs of leading
and higher twists at a low-momentum renormalization scale in a variety of chiral
quark models: the gauged non-local chiral quark model based on the instanton
picture of QCD vacuum \cite{Bowler:1994ir,Anikin:2000rq,Dorokhov:2003kf}, the
spectral quark model \cite{RuizArriola:2003bs}, and the Nambu--Jona-Lasinio
model with the Pauli-Villars regulator. There are a number of advantages of
the chiral model approach in comparison with other existing approaches.
Besides the incorporation of spontaneous chiral symmetry breaking calculations
can directly be carried out in the Minkowski space. Actually, the corresponding
calculations are manifestly covariant, so the subtleties of the light-cone
quantization do not arise. Within non-local models an effective resummation of
quark-gluon condensates of growing dimension is performed.

Although our results are obviously model dependent, we have theoretical
control on several properties that should be satisfied \textit{a priori}, with
the symmetry requirements being the most stringent restriction. The resulting
Ward-Takahashi identities are properly incorporated and their role in quark
models is very important when computing the light-cone properties. We stress that in our
approach we depart from standard parameterizations and do undertake a
\emph{genuine dynamical calculation}. Analogous results for the pion have been
obtained earlier in the instanton model
\cite{Esaibegian:1989uj,Anikin:1999cx,Dorokhov:2002iu}, in the
Nambu--Jona-Lasinio model \cite{RuizArriola:2002bp}, in the spectral quark
model \cite{RuizArriola:2003bs}, and most recently in the large-$N_{c}$ Regge
model \cite{Arriola:2006ii}.

The paper is organized as follows: in Sect. II we give definitions of the
photon DAs in different channels corresponding to the twist expansion up to the
twist-4 level. Necessary information about the chiral quark models that we are
going to use in the paper is given in Sect. III. In particular, the
corresponding normalization constants are evaluated. Section IV contains the
formalism of quark model calculations of the photon LCWF and DA and a proper
identification of the low-energy matrix elements is assessed. In Sect. V we give
the explicit expressions for the leading-twist photon and $\rho$-meson
distributions, while the perturbative QCD evolution of the resulting
leading-twist distributions to the relevant physical scales is carried out in
Sect. VI. In Section VII the calculation of the higher twist components of the
photon distributions is presented. Some technical details of
calculations, in particular the treatment of regularization and the QCD evolution, are given in
Appendices \ref{app:form} - \ref{app:QCD}.

\section{Definitions and notations}

The distribution amplitudes for the virtual photon are defined via the matrix
elements of quark-antiquark bilinear operators taken between the vacuum and
the one-photon state $|\gamma^{\lambda}(q)\rangle$ of momentum $q$ and
polarization vector $e_{\mu}^{\left(  \lambda\right)  }$ \footnote{Our
definitions and notation follow closely the works of Braun, Ball and coauthors
\cite{Ball:1998sk,Ball:2002ps}.}. It is assumed that the quark and antiquark
are separated by the distance $2z$ and the light-like limit $z^{2}%
\rightarrow0$ is taken at a fixed scalar product $q\cdot z$. We denote
$q^{2}>0$ for space-like vectors and $q^{2}<0$ for the time-like vectors. The
photon polarization is always perpendicular to $q$, thus we have
$e^{(\lambda)}\cdot q=0$, while for the case of the \emph{real} photon one has
in addition the condition $e^{(\lambda)}\cdot z=0$ constraining the photon
polarization to the two transverse directions.

Following Refs. \cite{Ali:1995uy,Ball:1996tb,Ball:1998sk,Ball:2002ps}, we use
the light-cone expansion of the matrix elements in order to define the
invariant amplitudes up to the twist-4 accuracy ($z^{2}$
terms are neglected)
\[
\langle0|\overline{q}(z)\sigma_{\mu\nu}[z,-z]q(-z)|\gamma^{\lambda}%
(q)\rangle=ie_{q}\left\langle 0\left\vert \overline{q}q\right\vert
0\right\rangle f_{\perp\gamma}^{t}\left(  q^{2}\right)  \left\{  \left(
e_{\mu}^{(\lambda)}q_{\nu}-q_{\mu}e_{\nu}^{(\lambda)}\right)  \chi
_{\mathrm{m}}\int_{0}^{1}dxe^{i\xi q\cdot z}\mathcal{A}_{T}(x,q^{2})+\right.
\]%
\begin{equation}
\left.  +\frac{e^{(\lambda)}\cdot z}{\left(  z\cdot q\right)  ^{2}}\left(
q_{\mu}z_{\nu}-z_{\mu}q_{\nu}\right)  \int_{0}^{1}dxe^{i\xi q\cdot
z}\mathcal{B}_{T}(x,q^{2})+\frac{1}{z\cdot q}\left(  e_{\mu}^{(\lambda)}%
z_{\nu}-z_{\mu}e_{\nu}^{(\lambda)}\right)  \int_{0}^{1}dxe^{i\xi q\cdot
z}\mathcal{C}_{T}(x,q^{2})\right\}  , \label{AT}%
\end{equation}%
\begin{align}
&  \langle0|\overline{q}(z)\gamma_{\mu}[z,-z]q(-z)|\gamma^{\lambda}%
(q)\rangle=e_{q}f_{3\gamma}f_{\perp\gamma}^{v}\left(  q^{2}\right)  \left\{
q_{\mu}\frac{e^{(\lambda)}\cdot z}{q\cdot z}\frac{f_{\parallel\gamma}%
^{v}\left(  q^{2}\right)  }{f_{\perp\gamma}^{v}\left(  q^{2}\right)  }\int
_{0}^{1}dxe^{i\xi q\cdot z}\mathcal{A}_{V}(x,q^{2})+\right. \nonumber\\
&  \left.  +\left(  e_{\mu}^{(\lambda)}-q_{\mu}\frac{e^{(\lambda)}\cdot
z}{q\cdot z}\right)  \int_{0}^{1}dxe^{i\xi q\cdot z}\mathcal{B}_{V}%
(x,q^{2})+z_{\mu}\frac{e^{(\lambda)}\cdot z}{\left(  q\cdot z\right)  ^{2}%
}\int_{0}^{1}dxe^{i\xi q\cdot z}\mathcal{C}_{V}(x,q^{2})\right\}
,\label{AV}\\
&  \langle0|\overline{q}(z)\gamma_{\mu}\gamma_{5}[z,-z]q(-z)|\gamma^{\lambda
}(q)\rangle=e_{q}f_{3\gamma}f_{\gamma}^{a}\left(  q^{2}\right)  \epsilon
_{\mu\nu\alpha\beta}e_{\nu}^{(\lambda)}q^{\alpha}z^{\beta}\int_{0}%
^{1}dxe^{i\xi q\cdot z}\mathcal{D}(x,q^{2}), \label{AAx}%
\end{align}
where $\left\langle 0\left\vert \overline{q}q\right\vert 0\right\rangle $ is
the quark condensate, $\chi_{\mathrm{m}}$ is the \emph{magnetic susceptibility
of the quark condensate}, and $f_{3\gamma}$ is related to the first moment of
the magnetic susceptibility. The symbol $[-z,z]$ in the matrix elements
denotes the path-ordered gauge link (Wilson line) for the gluon fields between
the points $-z$ and $z$. In the light-cone gauge, $A(z)\cdot z=0$, assumed
throughout our calculations, we have $[z,-z]=1$ and the Wilson lines may be
dropped. The integration variable $x$ corresponds to the momentum fraction
carried by the quark and $\xi=2x-1$ for the short-hand notation. The electric
charge of the quark is denoted by $e_{q}$. For a real photon, due to condition
$e^{(\lambda)}\cdot z=0$, only four structures corresponding to the invariant
amplitudes $\mathcal{A}_{T}$, $\mathcal{C}_{T}$, $\mathcal{B}_{V}$, and
$\mathcal{D}$ survive. In the present work we do not consider the twist-3
three-particle DAs which involve the gluon fields, integrated out in effective
quark models.

The corresponding decay constants and form factors are defined as%
\begin{align}
\langle0|\overline{q}(0)\sigma_{\mu\nu}q(0)|\gamma^{\lambda}(q)\rangle &
=ie_{q}\left\langle 0\left\vert \overline{q}q\right\vert 0\right\rangle
\chi_{\mathrm{m}}f_{\perp\gamma}^{t}\left(  q^{2}\right)  \left(  e_{\mu
}^{(\lambda)}q_{\nu}-q_{\mu}e_{\nu}^{(\lambda)}\right)  ,\label{ft}\\
\langle0|\overline{q}(0)\gamma_{\mu}q(0)|\gamma^{\lambda}(q)\rangle &
=e_{q}f_{3\gamma}f_{\perp\gamma}^{v}\left(  q^{2}\right)  e_{\mu}^{(\lambda
)},\label{fv}\\
\left.  \frac{\partial}{\partial z_{\beta}}\langle0|\overline{q}(z)\gamma
_{\mu}\gamma_{5}[z,-z]q(-z)|\gamma^{\lambda}(q)\rangle\right\vert
_{z\rightarrow0}  &  =e_{q}f_{3\gamma}f_{\gamma}^{a}\left(  q^{2}\right)
\epsilon_{\mu\nu\alpha\beta}e_{\nu}^{(\lambda)}q^{\alpha}. \label{fa}%
\end{align}
The vector constant at zero virtuality, $f_{\perp\gamma}^{v}\left(
q^{2}=0\right)  $, is zero due to the conservation of the vector current. The
nonperturbative constants $\chi_{\mathrm{m}}$ and $f_{3\gamma}$ provide
natural mass scales for the tensor and vector components of the photon DAs. In
QCD the matrix elements of the relevant local operators are defined as
\begin{align}
\left\langle 0\left\vert \overline{q}\sigma_{\mu\nu}q\right\vert
0\right\rangle _{F}  &  =e_{q}\chi_{m}\left\langle 0\left\vert \overline
{q}q\right\vert 0\right\rangle F_{\mu\nu},\\
\left\langle 0\left\vert \overline{q}g\widetilde{G}_{\mu\nu}\gamma_{\alpha
}\gamma_{5}q\right\vert 0\right\rangle _{F}  &  =e_{q}f_{3\gamma}D_{\alpha
}F_{\mu\nu},
\end{align}
where $F_{\mu\nu}$ $\left(  G_{\mu\nu}\right)  $ is the field-strength tensor
of the external electromagnetic (gluon) field, $D_{\alpha}$ is the covariant
derivative, and the index $F$ indicates that the vacuum expectation values are
taken in the QCD vacuum in the presence of the background field $F_{\mu\nu}$.
QCD predicts the scale dependence of the quark condensate $\left\langle
0\left\vert \overline{q}q\right\vert 0\right\rangle $, its magnetic
susceptibility $\chi_{\mathrm{m}}$, and $f_{3\gamma}$, which at the leading
logarithmic approximation evolve as%
\begin{equation}
\left.  \left\langle 0\left\vert \overline{q}q\right\vert 0\right\rangle
\right\vert _{\mu}=L^{-\gamma_{\overline{q}q}/b}\left.  \left\langle
0\left\vert \overline{q}q\right\vert 0\right\rangle \right\vert _{\mu_{0}%
},\quad\left.  \chi_{m}\right\vert _{\mu}=L^{-\left(  \gamma_{0}%
-\gamma_{\overline{q}q}\right)  /b}\left.  \chi_{m}\right\vert _{\mu_{0}%
},\quad\left.  f_{3\gamma}\right\vert _{\mu}=L^{-\gamma_{f}/b}\left.
f_{3\gamma}\right\vert _{\mu_{0}} \label{scale}%
\end{equation}
where $L=\alpha_{s}\left(  \mu^{2}\right)  /\alpha_{s}\left(  \mu_{0}%
^{2}\right)  $ is the evolution ratio, $b=\left(  11N_{c}-2n_{f}\right)  /3$,
$\gamma_{\overline{q}q}=-3C_{F}$ is the anomalous dimension of the quark
condensate, $\gamma_{0}=C_{F}$ is the anomalous dimension of the chiral-odd
local operator of leading-twist, $\gamma_{f}=3C_{A}-C_{F}/3$, with $C_{F}=4/3$
and $C_{A}=3$ for $N_{c}=3$. By convention, the constants are chosen in such a
way that the invariant functions are normalized by conditions%
\begin{align}
\int_{0}^{1}dx\mathcal{A}_{T}(x,q^{2})  &  =1,\qquad\int_{0}^{1}%
dx\mathcal{B}_{T}(x,q^{2})=\int_{0}^{1}dx\mathcal{C}_{T}(x,q^{2})=0,\\
\int_{0}^{1}dx\mathcal{A}_{V}(x,q^{2})  &  =\int_{0}^{1}dx\mathcal{B}%
_{V}(x,q^{2})=1,\qquad\int_{0}^{1}dx\mathcal{C}_{V}(x,q^{2})=0,\nonumber\\
\int_{0}^{1}dx\mathcal{D}(x,q^{2})  &  =1.\nonumber
\end{align}

In order to define the photon DAs of definite twist it is convenient to
introduce the light-like vector $p_{\mu}$ $\left(  p^{2}=0\right)  $, such
that%
\begin{equation}
p_{\mu}=q_{\mu}-\frac{q^{2}}{2}n_{\mu},
\end{equation}
where the light-like vector%
\begin{equation}
n_{\mu}=\frac{z_{\mu}}{p\cdot z}%
\end{equation}
is normalized by the condition $n\cdot p=1$. Then the photon polarization
vector $e_{\mu}^{(\lambda)}$ is decomposed into projections onto the two
light-like vectors and the transverse plane as%
\begin{equation}
e_{\mu}^{(\lambda)}=\left(  e^{(\lambda)}\cdot n\right)  p_{\mu}+\left(
e^{(\lambda)}\cdot p\right)  n_{\mu}+e_{\perp\mu}^{(\lambda)}.
\end{equation}
Applying the useful relations%
\begin{equation}
z\cdot q=z\cdot p,\qquad e^{(\lambda)}\cdot p=-\frac{q^{2}}{2}\left(
e^{(\lambda)}\cdot n\right)  ,
\end{equation}
it is possible to express the above definitions in the form
\begin{align}
&  \langle0|\overline{q}(z)\sigma_{\mu\nu}q(-z)|\gamma^{\lambda}%
(q)\rangle=ie_{q}\langle\bar{q}q\rangle f_{\perp\gamma}^{t}\left(
q^{2}\right)  \left\{  \left(  e_{\perp\mu}^{(\lambda)}p_{\nu}-e_{\perp\nu
}^{(\lambda)}p_{\mu}\right)  \chi_{\mathrm{m}}\int_{0}^{1}dxe^{i\xi q\cdot
z}\phi_{\perp\gamma}(x,q^{2})+\right. \nonumber\\
&  +\left.  \left(  p_{\mu}n_{\nu}-n_{\mu}p_{\nu}\right)  \left(
e^{(\lambda)}n\right)  \int_{0}^{1}dxe^{i\xi q\cdot z}\psi_{\gamma}^{\left(
t\right)  }(x,q^{2})+\left(  e_{\perp\mu}^{(\lambda)}n_{\nu}-n_{\mu}%
e_{\perp\nu}^{(\lambda)}\right)  \int_{0}^{1}dxe^{i\xi q\cdot z}h_{\gamma
}^{\left(  t\right)  }(x,q^{2})\right\}  , \label{phi}%
\end{align}%
\begin{align}
&  \langle0|\overline{q}(z)\gamma_{\mu}q(-z)|\gamma^{\lambda}(q)\rangle
=e_{q}f_{3\gamma}f_{\perp\gamma}^{v}\left(  q^{2}\right)  \left\{  p_{\mu
}\left(  e^{(\lambda)}\cdot n\right)  \frac{f_{\parallel\gamma}^{v}\left(
q^{2}\right)  }{f_{\perp\gamma}^{v}\left(  q^{2}\right)  }\int_{0}%
^{1}dxe^{i\xi q\cdot z}\phi_{\parallel\gamma}(x,q^{2})+\right. \nonumber\\
&  \left.  +e_{\perp\mu}^{(\lambda)}\int_{0}^{1}dxe^{i\xi q\cdot z}\psi
_{\perp\gamma}^{(v)}(x,q^{2})+n_{\mu}\left(  e^{(\lambda)}\cdot n\right)
\int_{0}^{1}dxe^{i\xi q\cdot z}h_{\gamma}^{(v)}(x,q^{2})\right\}
,\label{phiV}\\
&  \langle0|\overline{q}(z)\gamma_{\mu}\gamma_{5}q(-z)|\gamma^{\lambda
}(q)\rangle=e_{q}f_{3\gamma}f_{\gamma}^{a}\left(  q^{2}\right)  \epsilon
_{\mu\nu\alpha\beta}e_{\nu}^{(\lambda)}p^{\alpha}z^{\beta}\int_{0}%
^{1}dxe^{i\xi q\cdot z}\psi_{\gamma}^{(a)}(x,q^{2}). \label{phiA}%
\end{align}
The DAs $\phi_{\perp\gamma}$ and $\phi_{\parallel\gamma}$ are of twist-2,
$\psi_{\gamma}^{\left(  t,v,a\right)  }$ are of twist-3, and $h_{\gamma
}^{\left(  t,v\right)  }$ are of twist-4\footnote{This dimensional
twist-counting refers to the \textquotedblleft dynamical\textquotedblright%
\ twist of a matrix-element, as opposed to the \textquotedblleft
geometric\textquotedblright\ twist of a (local) operator.}. In the case of the
real photon one has $e^{(\lambda)}\cdot n=0$ and only $\phi_{\perp\gamma}$,
$\psi_{\perp\gamma}^{(v)}$, $\psi_{\gamma}^{(a)}$ and $h_{\gamma}^{\left(
t\right)  }$ are coupled. The leading twist functions $\phi_{\perp\gamma
}(x,q^{2})$ and $\phi_{\parallel\gamma}(x,q^{2})$ describe the photon
distribution amplitudes that a virtual photon with momentum $q$ dissociates
into a quark-antiquark pair at small transverse separation.

Comparing (\ref{phi})-(\ref{phiA}) with the light-cone expansions
(\ref{AT})-(\ref{AAx}) one finds the following relations between the invariant amplitudes
and the DAs%
\begin{align}
\phi_{\perp\gamma}\left(  x,q^{2}\right)   &  =\mathcal{A}_{T}\left(
x,q^{2}\right)  ,\quad\psi_{\gamma}^{\left(  t\right)  }\left(  x,q^{2}%
\right)  =q^{2}\chi_{\mathrm{m}}\mathcal{A}_{T}\left(  x,q^{2}\right)
+\mathcal{B}_{T}\left(  x,q^{2}\right)  +\mathcal{C}_{T}\left(  x,q^{2}%
\right)  ,\quad\nonumber\\
h_{\gamma}^{\left(  t\right)  }\left(  x,q^{2}\right)   &  =\frac{1}{2}%
q^{2}\chi_{\mathrm{m}}\mathcal{A}_{T}\left(  x,q^{2}\right)  +\mathcal{C}%
_{T}\left(  x,q^{2}\right)  ,\nonumber\\
\phi_{\parallel\gamma}\left(  x,q^{2}\right)   &  =\mathcal{A}_{V}\left(
x,q^{2}\right)  ,\quad\psi_{\gamma}^{\left(  v\right)  }\left(  x,q^{2}%
\right)  =\mathcal{B}_{V}\left(  x,q^{2}\right)  ,\quad\nonumber\\
h_{\gamma}^{\left(  v\right)  }\left(  x,q^{2}\right)   &  =\frac{1}{2}%
q^{2}\frac{f_{\parallel\gamma}^{v}\left(  q^{2}\right)  }{f_{\perp\gamma}%
^{v}\left(  q^{2}\right)  }\mathcal{A}_{V}\left(  x,q^{2}\right)
-q^{2}\mathcal{B}_{V}\left(  x,q^{2}\right)  +\mathcal{C}_{V}\left(
x,q^{2}\right)  ,\quad\nonumber\\
\psi_{\gamma}^{\left(  a\right)  }\left(  x,q^{2}\right)   &  =\mathcal{D}%
\left(  x,q^{2}\right)  .
\end{align}
Within the quark models it is more convenient to calculate first the invariant
amplitudes and then by using the above relations the DAs themselves.

\section{Chiral quark models}

\subsection{ Instanton-motivated nonlocal chiral quark model}

The analysis in chiral quark models is carried at the one-quark-loop level,
\emph{i.e.} at leading order in the number of colors, $N_{c}$. We
consider the strict chiral limit assuming zero current quark masses,
$m_{u,d}=0$. Our basic expressions are derived for the case of the
\emph{non-local chiral quark model}, discussed in detail
Ref.~\cite{Terning:1991yt,Holdom:1992fn,Plant:1997jr,Broniowski:1999bz,Anikin:2000rq}%
. The expressions for other models analyzed in this paper can be formally
obtained from the general formulas of the non-local model.

The non-local quark models are inspired by the underlying QCD-based models,
such as the instanton-vacuum model
\cite{Shuryak:1981ff,Diakonov:1983hh,Kochelev:1985de,Dorokhov:1992hk}, the
Schwinger-Dyson resummation of the rainbow diagrams \cite{Roberts:1994dr}, and
some others, which all may be cast in the form of non-local quark dynamics. The
models develop, via spontaneous chiral symmetry breaking, the quark mass
depending on the quark virtuality. The quark propagator has the form
(following the instanton liquid model we set for simplicity the quark
wave-function renormalization to unity, $Z(p)=1$)%
\begin{equation}
S(p)=\frac{1}{\widehat{p}-M(p)+i\varepsilon},\qquad M(p)=M_{0}f^{2},
(p^{2})\label{prop}%
\end{equation}
where the dynamical quark mass $M(p)$ is expressed via the nonlocal (in coordinate space)
function $f\left(  p\right)  $ defining the nonlocal properties of the QCD vacuum
\cite{Mikhailov:1991pt,Dorokhov:1997iv,Dorokhov:1999ig}. We have to emphasize
the essential difference of this expression compared to the usually used
perturbative expression, where the effects of the spontaneous violation of the
chiral symmetry are not taken into account and thus $M(p)\equiv0$.

Throughout the paper we use the notations\footnote{In following the integrals
over the momentum are calculated by transforming the integration variables
into the Euclidean space, ($k^{0}\rightarrow ik_{4},$ $k^{2}\rightarrow-k^{2}%
$).}%
\begin{align}
D(k)  &  =k^{2}+M\left(  k\right)  ,\qquad\qquad M^{(1)}\left(  k,k^{\prime
}\right)  =\frac{M\left(  k\right)  -M\left(  k^{\prime}\right)  }%
{k^{2}-k^{\prime2}},\\
k_{\pm}  &  =k\pm\frac{1}{2}q,\qquad k_{\perp}^{2}=\left(  k_{+}\cdot
k_{-}\right)  -\frac{\left(  k_{+}\cdot q\right)  \left(  k_{-}\cdot q\right)
}{q^{2}},\qquad
\end{align}
and for any function $F$ one defines $F_{^{\pm}}=F\left(  k_{\pm}\right)  .$

The photon-quark coupling for the incoming quark of momentum $k$ and the
outgoing quark of momentum $k^{\prime}=k+q$ is equal to
\cite{Plant:1997jr,Dorokhov:2005pg}
\begin{equation}
\Gamma^{\mu}(k,k^{\prime})=\gamma_{\mu}-(k+k^{\prime})_{\mu}M_{k,k^{\prime}%
}^{(1)}+\left(  \gamma_{\mu}-\frac{q_{\mu}\widehat{q}}{q^{2}}\right)
B_{V}\left(  q^{2}\right)  , \label{vert1}%
\end{equation}
where the contribution of the rescattering in the $\rho$ meson channel is
taken into account by the factor \cite{Dorokhov:2005pg}
\begin{equation}
B_{V}\left(  q^{2}\right)  =\frac{G_{V}}{1-G_{V}J_{V}^{T}\left(  q^{2}\right)
}8N_{c}\int\frac{d^{4}k}{\left(  2\pi\right)  ^{4}}\frac{f_{+}^{V}f_{-}^{V}%
}{D_{+}D_{-}}\left[  M_{+}M_{-}+k_{+}\cdot k_{-}-\frac{2}{3}k_{\perp}%
^{2}\left(  1+M^{2(1)}\left(  k_{+},k_{-}\right)  \right)  +\frac{4}%
{3}k_{\perp}^{2}\frac{f_{-}f^{(1)}(k_{-},k_{+})}{D_{-}}\right]  . \label{BV}%
\end{equation}
The polarization operator in the vector channel is given by
\begin{equation}
J_{V}^{T}(q^{2})=-2N_{c}\int\frac{d^{4}k}{\left(  2\pi\right)  ^{4}}%
\frac{\left(  f_{+}^{V}f_{-}^{V}\right)  ^{2}}{D_{+}D_{-}}\left[  M_{+}%
M_{-}+k_{+}\cdot k_{-}-\frac{2}{3}k_{\perp}^{2}\right]  . \label{J}%
\end{equation}
Importantly, the vertex (\ref{vert1}) satisfies the Ward-Takahashi identity
for the vector current,
\begin{equation}
q_{\mu}\Gamma^{\mu}(k,k^{\prime})=S^{-1}%
(k)-S^{-1}(k^{\prime}).
\end{equation}
The straightforward check uses the fact that
$B_{V}\left(  0\right)  =0$ identically. The inclusion of dressed vertices is
crucial for the consistency of calculations in the non-local quark models.

As argued in \cite{Dorokhov:2005pg} (see also \cite{Dorokhov:1997iv,Dorokhov:1998up}), it is
convenient to work with the mass function corresponding to the instanton field
taken in the axial gauge, where $P$exponential factor over nonperturbative (instanton) gluon
field equals unity. In this gauge at large space-like momenta $M(p)$
must decrease faster than any inverse power of $p^{2}$, \emph{e.g.}, as an
exponential. Below we assume for simplicity a universal gaussian form of the
two nonlocal functions,
\begin{equation}
f(p)=f_{V}\left(  p\right)  =\exp\left(  -\frac{p^{2}}{\Lambda^{2}}\right)
,\label{GaussFF}%
\end{equation}
with $p$ denoting the Euclidean momentum. Also in this work we do not
consider for the nonlocal model the time-like region close to the resonances,
such as the $\rho$-meson. As the model parameters we take
\begin{equation}
M_{0}=240~\mathrm{MeV},\;\;\Lambda=1110~\mathrm{MeV},\label{nlparam}%
\end{equation}
from the work \cite{Dorokhov:2004ze} where they were fixed by the requirement
to reproduce the chiral limit of the pion weak decay constant $f_{\pi
}^{\mathrm{chiral}}=86$ MeV \cite{Gasser:1983yg} and the contribution of the
hadronic vacuum polarization to the anomalous magnetic moment of the muon. As
found in \cite{Dorokhov:2004ze}, the latter is very sensitive to the value of
the mass parameter $M_{0}$, preferring lower numbers. For the vector coupling
we take $G_{V}=-8.7\quad\mathrm{GeV}^{-2}.$ In Appendix \ref{app:Gauss} we give
further details on the properties of the Gaussian shape of nonlocality in momentum
and $\alpha$-representations.

Let us also mention restriction on the nonlocal models coming from consistency
with the operator product expansion in QCD. It was first found in
\cite{Dorokhov:2002iu} that in order to guarantee correct high momenta
expansion of hadronic form factors and other quantities only vertices
depending on quark virtualities $k^{2}$ and $k^{\prime2}$ may be used. The possible forms are
$f(k^{2})f\left(  k^{\prime2}\right)  $, as it is predicted by instanton
model, or more general form $f(\frac{k^{2}+k^{\prime2}}{2})$. The vertices containing
the linear combination of quark momenta, such as $f(\frac{k+k^{\prime}}{2}),$
lead to exponentially growing hadronic form factors at large external momenta.
The same requirement restricts possible prescriptions for the usage of the
flavor $P$-exponents gauged with respect to external currents.
% nonlocal effective interaction.
We follow the prescription suggested by Mandelstam and
first used within the effective models in \cite{Terning:1991yt}. On the other
hand, the prescription used in \cite{Bowler:1994ir} leads to a linear
combination of quark momenta and results in exponentially growing
asymptotics of hadronic form factors, inconsistent with the operator product expansion.

Now we are ready to compute the quantities referring to the quark bilinear
matrix elements of Eq.~(\ref{AT}-\ref{AAx}). The normalization factors are
defined as (we denote $u=k^{2}$)\footnote{The expression for the magnetic
susceptibility has been obtained in \cite{Kim:2004hd,Dorokhov:2005pg}.}
\begin{equation}
\left\langle 0\left\vert \overline{q}q\right\vert 0\right\rangle
^{\mathrm{inst}}=-N_{c}\int\frac{du}{4\pi^{2}}\frac{uM(u)}{D\left(  u\right)
}, \label{QQI}%
\end{equation}%
\begin{equation}
\chi_{\mathrm{m}}^{\mathrm{inst}}=-\frac{N_{c}}{\langle\bar{q}q\rangle}%
\int\frac{du}{4\pi^{2}}\frac{u\left(  M\left(  u\right)  -uM^{\prime}\left(
u\right)  \right)  }{D^{2}\left(  u\right)  }, \label{ChiM}%
\end{equation}%
\begin{equation}
f_{3\gamma}^{\mathrm{inst}}=-N_{c}\int\frac{du}{4\pi^{2}}\frac{M^{2}\left(
u\right)  }{D\left(  u\right)  }. \label{f3gam}%
\end{equation}
The above equations with parameters (\ref{nlparam}) yield%
\begin{align}
\left.  \left\langle 0\left\vert \overline{q}q\right\vert 0\right\rangle
^{\mathrm{inst}}\right\vert _{\mu_{\mathrm{inst}}}  &  =-(0.214~\mathrm{GeV)}%
^{3},\quad\left.  \chi_{\mathrm{m}}^{\mathrm{inst}}\right\vert _{\mu
_{\mathrm{inst}}}=4.32\quad\mathrm{GeV}^{-2},\label{f3g_Inst}\\
\left.  f_{3\gamma}^{\mathrm{inst}}\right\vert _{\mu_{\mathrm{inst}}}  &
=-0.0073\quad\mathrm{GeV}^{2}.\nonumber
\end{align}
In Ref. \cite{Dorokhov:2005pg} the renormalization scale typical for the
instanton fluctuations was estimated as $\mu_{\mathrm{inst}}\approx0.53$~GeV.
The estimate is made by means of comparing the value of the quark condensate with
that found in QCD sum rules at standard renormalization point $\mu
_{0}=1~\mathrm{GeV}$. With $\Lambda_{\mathrm{QCD}}$ for three flavors being
$\Lambda_{\mathrm{QCD}}^{\left(  n_{f}=3\right)  }=312$~MeV
\cite{Kataev:2001kk,Bakulev:2002uc} the evolution ratio in (\ref{scale}) from
the scale $\mu_{\mathrm{inst}}$ to $\mu_{0}$ equals $L=2.17$. Rescaling the
values obtained in the non-local quark model to the standard renormalization
point $\mu_{0}$ yields
\begin{align}
\left.  \left\langle 0\left\vert \overline{q}q\right\vert 0\right\rangle
^{\mathrm{inst}}\right\vert _{\mu_{0}}  &  =-(0.24~\mathrm{GeV)}^{3}%
,\quad\left.  \chi_{\mathrm{m}}^{\mathrm{inst}}\right\vert _{\mu_{0}%
}=2.73\quad\mathrm{GeV}^{-2},\\
\left.  f_{3\gamma}^{\mathrm{inst}}\right\vert _{\mu_{0}}  &  =-0.0035\quad
\mathrm{GeV}^{2}.\nonumber
\end{align}
These numbers are in a rather good agreement with the estimates obtained from
the QCD sum rules supplemented by the vector meson dominance technique, which
provide \cite{Ball:2002ps}%
\begin{align}
\left.  \left\langle 0\left\vert \overline{q}q\right\vert 0\right\rangle
^{\mathrm{QCDsr}}\right\vert _{\mu_{0}}  &  =-(0.24\pm0.02)~\mathrm{GeV}%
^{3},\quad\left.  \chi_{\mathrm{m}}^{\mathrm{QCDsr}}\right\vert _{\mu_{0}%
}=(3.15\pm0.3)\quad\mathrm{GeV}^{-2},\label{f3g_QCD}\\
\left.  f_{3\gamma}^{\mathrm{QCDsr}}\right\vert _{\mu_{0}}  &  =-(0.0039\pm
0.0020)\quad\mathrm{GeV}^{2}.\nonumber
\end{align}
Let us also mention that the vector-meson dominance model
predicts~\cite{Ball:2002ps}
\begin{equation}
\chi_{\mathrm{m}}^{\mathrm{vmd}}=\frac{2}{m_{\rho}^{2}}=3.37~\mathrm{GeV}%
^{-2},\quad f_{3\gamma}^{\mathrm{vmd}}=-f_{\rho}^{2}\varsigma_{3\rho
}=-(0.0046\pm0.0020)\quad\mathrm{GeV}^{2}, \label{f3g_VMD}%
\end{equation}
where $m_{\rho}=770$ MeV, $f_{\rho}=215$ MeV, and $\varsigma_{3\rho}%
=0.10\pm0.05$.

\begin{figure}[h]
\begin{center}
\includegraphics[width=5cm]{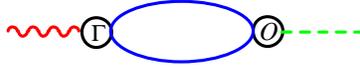}
\end{center}
\caption{{Diagrammatic representation of the photon-to-current transition. }}%
\label{diag}%
\end{figure}

\begin{figure}[h]
\begin{center}
\includegraphics[width=9cm]{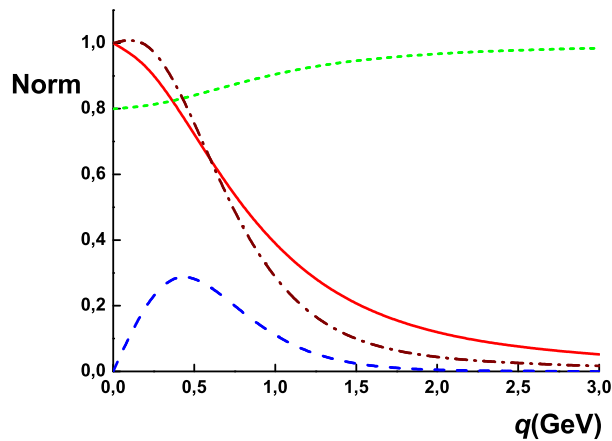}
\end{center}
\caption{Form factors associated with the photon distribution amplitude
plotted as functions of the space-like momentum squared (solid line for the
tensor form factor $f_{\perp\gamma}^{t}$, dash-dot line for the vector
longitudinal form factor $f_{\parallel\gamma}^{v}$, dashed line for the vector
transverse form factor $f_{\perp\gamma}^{v}$, and short-dashed line for the
axial form factor $f_{\gamma}^{a}$). }%
\label{Norm}%
\end{figure}

The calculations of the form factors (\ref{ft})-(\ref{fa}) consist of the
evaluation of the one-loop diagram of Fig.~\ref{diag}, where one vertex is the
photon-quark coupling $e_{q}\Gamma\left(  k,k^{\prime}\right)  \cdot
e^{\left(  \lambda\right)  }\left(  q\right)  $, and the other vertex involves
the operators $\sigma_{\alpha\beta}$, $\gamma_{\mu}$, or $\gamma_{\mu}%
\gamma_{5}$. The photon DAs, discussed later on, are obtained from these
one-loop diagrams with the light-cone quark momentum projection fixed. The
derivation of the form factors is straightforward \cite{Dorokhov:2005pg} and
yields
\begin{equation}
f_{\perp\gamma}^{t}\left(  q^{2}\right)  =-\frac{N_{c}}{\chi_{\mathrm{m}%
}\left\langle 0\left\vert \overline{q}q\right\vert 0\right\rangle }\int
\frac{d^{4}k}{4\pi^{4}}\frac{1}{D_{+}D_{-}}\left\{  \left[  \frac{M_{+}+M_{-}%
}{2}-\frac{k\cdot q}{q^{2}}\left(  M_{+}-M_{-}\right)  \right]  \left(
1+B_{V}\left(  q^{2}\right)  f_{+}^{V}f_{-}^{V}\right)  -\frac{2}{3}k_{\perp
}^{2}M^{(1)}(k_{+},k_{-})\right\}  , \label{ChiQ}%
\end{equation}%
\begin{equation}
f_{\parallel\gamma}^{v}\left(  q^{2}\right)  =-\frac{N_{c}}{f_{3\gamma}}%
\int\frac{d^{4}k}{4\pi^{4}}\frac{1}{D_{+}D_{-}}\left[  \left(  M_{+}%
M_{-}+\left[  k_{+}\cdot k_{-}-\frac{4}{3}k_{\perp}^{2}\right]  _{\mathrm{sub}%
}\right)  \left(  1+B_{V}\left(  q^{2}\right)  f_{+}^{V}f_{-}^{V}\right)
-\frac{4}{3}k_{\perp}^{2}M^{^{2}(1)}(k_{+},k_{-})\right],
\end{equation}%
\begin{equation}
f_{\perp\gamma}^{v}\left(  q^{2}\right)  =-\frac{N_{c}}{f_{3\gamma}}\int
\frac{d^{4}k}{4\pi^{4}}\frac{1}{D_{+}D_{-}}\left[  \left(  M_{+}M_{-}+\left[
k_{+}\cdot k_{-}-\frac{2}{3}k_{\perp}^{2}\right]  _{\mathrm{sub}}\right)
\left(  1+B_{V}\left(  q^{2}\right)  f_{+}^{V}f_{-}^{V}\right)  -\frac{2}%
{3}k_{\perp}^{2}M^{^{2}(1)}(k_{+},k_{-})\right]  ,
\end{equation}
\begin{equation}
f_{\gamma}^{a}\left(  q^{2}\right)  =1-\frac{1N_{c}}{3f_{3\gamma}}\int
\frac{d^{4}k}{4\pi^{4}}k_{\perp}^{2}\left[  \frac{M_{+}^{2}-M_{-}^{2}%
+M_{-}^{2\prime}\left(  u_{-}+M_{+}^{2}\right)  }{D_{+}D_{-}^{2}}+\left(
+\leftrightarrow-\right)  \right]
\end{equation}
The subscript $\emph{sub}$ means that one needs to subtract the perturbative
contribution corresponding to the same expression with $M(k)$ set to zero. The
results are presented in Fig.~\ref{Norm}. We note that except for $f_{\gamma
}^{a}$, the form factors drop with the characteristic scale of about 1~GeV.

In the case of the real photon the one-loop calculations provide the
normalization constants%
\begin{align}
f_{\perp\gamma}^{t}\left(  0\right)   &  =1,\qquad f_{\perp\gamma}^{v}\left(
q^2\right)  =O(q^{2}),\\
f_{\parallel\gamma}^{v}\left(  0\right)   &  =1,\qquad f_{\gamma}^{a}\left(
0\right)  =1-\frac{N_{c}}{f_{3\gamma}}\int\frac{du}{8\pi^{2}}\frac
{M^{4}\left(  u\right)  }{D^{2}\left(  u\right)  }.\nonumber
\end{align}
The vanishing of the transverse vector part at zero $q^{2}$ is just the
manifestation of the Ward-Takahashi identity. The axial constant is not zero
because the spontaneous breaking of chiral symmetry is taken into account.

At large Euclidean $q$ one has%
\begin{align}
f_{\perp\gamma}^{t}\left(  q\rightarrow\infty\right)   &  =\frac{2}{\chi_{m}%
}\frac{1}{q^{2}},\qquad f_{\perp\gamma}^{v}\left(  q\rightarrow\infty\right)
=O(q^{-4}),\label{ChiAs}\\
f_{\parallel\gamma}^{v}\left(  q\rightarrow\infty\right)   &  =-\frac{N_{c}%
}{f_{3\gamma}}\int\frac{du}{4\pi^{2}}\frac{uM^{2}\left(  u\right)  }{D\left(
u\right)  }\frac{1}{q^{2}},\qquad f_{\gamma}^{a}\left(  q\rightarrow
\infty\right)  =1.\nonumber
\end{align}

\subsection{Spectral quark model}

\begin{figure}[tb]
\begin{center}
\includegraphics[width=7cm]{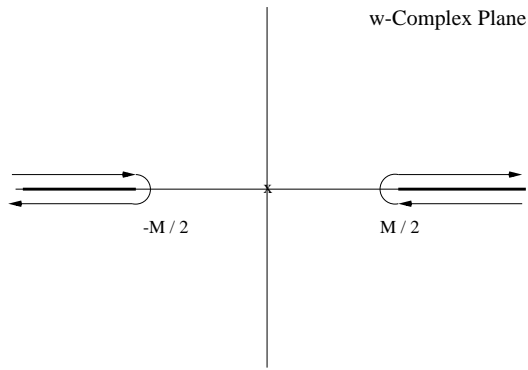}
\end{center}
\caption{The contour $C$ for the $\omega$-integration in the spectral quark
model. $M$ denotes the $\rho$-meson mass.}%
\label{contour}%
\end{figure}

The spectral quark model (SQM) is based on the spectral representation of the
quark propagator \cite{RuizArriola:2003bs}. The expressions for one-loop
observables in SQM are obtained from the expressions of the preceding sections
by replacing the quark mass with the spectral mass, $M\rightarrow\omega$, and
then integrating over omega with the spectral density $\rho(\omega)$. We have
\[
A=\int_{C}d\omega\rho(\omega)A(\omega),
\]
where $C$ is a suitably chosen contour in the complex $\omega$ plane. In the
meson-dominance version of SQM \cite{RuizArriola:2003bs} we have
\begin{align}
\rho(\omega)  &  =\rho_{V}(\omega)+\rho_{S}(\omega),\\
\rho_{V}(\omega)  &  =\frac{1}{2\pi i}\frac{1}{\omega}\frac{1}{(1-4\omega
^{2}/M_{V}^{2})^{5/2}},\nonumber\\
\rho_{S}(\omega)  &  =-\frac{1}{2\pi iN_{c}M_{S}^{4}}\frac{48\pi^{2}%
\langle0\left\vert \bar{q}q\right\vert 0\rangle}{(1-4\omega^{2}/M_{S}%
^{2})^{5/2}}, \label{mesdom}%
\end{align}
where $M_{V}$ is the mass of the $\rho$ meson, and $m_{S}$ provides the scale
for the scalar spectral density. We note that the quark condensate is a model
parameter. Matching the SQM predictions to the large-$N_{c}$ results of the chiral
perturbation theory in the single resonance exchange approximation allows for
the identification \cite{Megias:2004uj}
\[
M_{S}=M_{V}=m_{\rho},
\]
where $m_{\rho}$ is the mass of the $\rho$ meson. The contour $C$ for the
integration over the spectral mass $\omega$ is displayed in Fig.~\ref{contour}.

SQM with the meson dominance (\ref{mesdom}) generates, by construction, the
monopole form of the pion electromagnetic form factor
\cite{RuizArriola:2003bs}. Interestingly, the model has the feature of the
analytic quark confinement \emph{ }\cite{Efimov:1993zg}, \emph{i.e.} the quark
propagator has no poles, only cuts, in the complex momentum plane. Moreover,
the evaluation of low-energy matrix elements in SQM is very simple and leads
to numerous results reported in Ref.~\cite{RuizArriola:2003bs}, in particular
for the pion light-cone wave function and the pion structure functions.

The results for the normalization factors in SQM are
\[
\left.  \chi_{\mathrm{m}}^{\mathrm{SQM}}\right\vert _{\mu_{\mathrm{SQM}}%
}=\frac{2}{m_{\rho}^{2}}=3.37~\mathrm{GeV}^{-2},\;\;\left.  f_{3\gamma
}^{\mathrm{SQM}}\right\vert _{\mu_{\mathrm{SQM}}}=-\frac{N_{c}}{4\pi^{2}}%
\frac{m_{\rho}^{2}}{6}=-0.0075~\mathrm{GeV}^{2}.
\]
In Ref.~\cite{RuizArriola:2003bs} it has been argued that the quark model
scale corresponding to SQM is very low, $\mu_{\mathrm{SQM}}=313$~MeV. That
yields a large evolution ratio $L=4.6$ between $\mu_{\mathrm{SQM}}$ and
$\mu_{\mathrm{0}}=$1~GeV, and, consequently, we find after evolution
\[
\left.  \chi_{\mathrm{m}}^{\mathrm{SQM}}\right\vert _{\mu_{\mathrm{0}}%
}=1.37~\mathrm{GeV}^{-2},\;\;\left.  f_{3\gamma}^{\mathrm{SQM}}\right\vert
_{\mu_{\mathrm{0}}}=-0.0018~\mathrm{GeV}^{2},
\]
much lower values than in the nonlocal model and QCD\ sum rule.

\begin{figure}[h]
\begin{center}
\includegraphics[width=9cm]{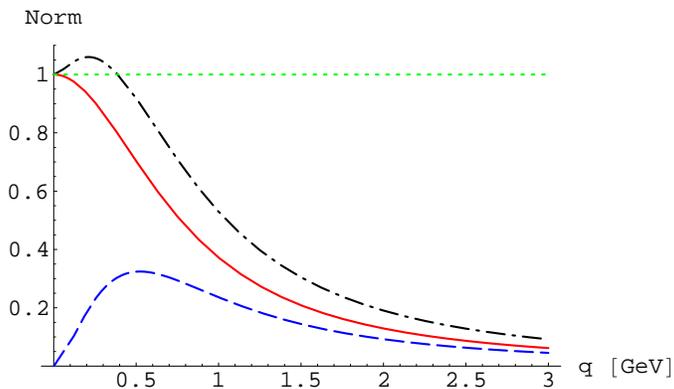}
\end{center}
\caption{Same as Fig.~\ref{Norm} for the spectral quark model.}%
\label{Normwb}%
\end{figure}

For the tensor form factor we find
\[
f_{\perp\gamma}^{t}\left(  q^{2}\right)  =\frac{m_{\rho}^{2}}{m_{\rho}%
^{2}+q^{2}},
\]
in full accordance to the vector dominance model. This makes a strong case
for the assumed scalar part of the meson-dominance of spectral quark model,
with the power $d_{S}=5/2$ (see Ref.~\cite{RuizArriola:2003bs}). The tensor
form factor satisfies the conditions
\begin{equation}
f_{\perp\gamma}^{t}\left(  0\right)  =1,\;\;f_{\perp\gamma}^{t}\left(
Q^{2}\rightarrow\infty\right)  =\frac{2}{\chi_{\mathrm{m}}q^{2}}. \label{48}%
\end{equation}
For the axial form factor we find in local models $f^{a}_{\gamma}(q^{2})=1$.
Expressions for the vector form factors, $f^{v}_{\gamma\parallel}$ and
$f^{v}_{\gamma\perp}$ are, due to the vacuum subtraction, not analytic and the
integration over the $x$ variable has to be carried out numerically. We
present these form factors for SQM in Fig.~\ref{Normwb}, noting the same
qualitative behavior as for the non-local model in Fig.~\ref{Norm}.

\subsection{Nambu--Jona-Lasinio model}

The expressions for one-loop observables in the Nambu--Jona-Lasinio model
(NJL) with the Pauli-Villars regulator follow from the expressions for the
non-local model. Formally, one replaces in the denominator the square of the
(constant) mass, $M^{2}\rightarrow M^{2}+\Lambda^{2}$. Then, in the simplest
twice subtracted case one has for any observable $A$ the prescription
\[
A=A(\Lambda^{2}=0)-A(\Lambda^{2})+\Lambda^{2}\frac{dA(\Lambda^{2})}%
{d\Lambda^{2}}.
\]
The parameters used are $M=280$~MeV and $\Lambda=871$~MeV, which were adjusted
to yield $F_{\pi}=93$~MeV. The used NJL model has no explicit coupling to
vector mesons ($G_{V}=0$). For the normalization factors we get%

\begin{align}
\left.  \left\langle 0\left\vert \overline{q}q\right\vert 0\right\rangle
\right\vert _{\mu_{\mathrm{NJL}}}  &  =\frac{N_{c}M}{4\pi^{2}}\left[
\Lambda^{2}-M^{2}\log\left(  1+\frac{\Lambda^{2}}{M^{2}}\right)  \right]
=-(0.230~\mathrm{GeV)}^{3},\nonumber\\
\quad\left.  \chi_{\mathrm{m}}\right\vert _{\mu_{\mathrm{NJL}}}  &
=-\frac{N_{c}M}{4\pi^{2}\left\langle 0\left\vert \overline{q}q\right\vert
0\right\rangle }\left[  \frac{\Lambda^{2}}{M^{2}+\Lambda^{2}}-\log\left(
1+\frac{\Lambda^{2}}{M^{2}}\right)  \right]  =2.55\quad\mathrm{GeV}%
^{-2},\nonumber\\
\left.  f_{3\gamma}\right\vert _{\mu_{\mathrm{NJL}}}  &  =\frac{N_{c}M^{2}%
}{4\pi^{2}}\left[  \frac{\Lambda^{2}}{M^{2}+\Lambda^{2}}-\log\left(
1+\frac{\Lambda^{2}}{M^{2}}\right)  \right]  =-0.0087~\mathrm{GeV}^{2}.
\label{f3g_NJL}%
\end{align}
The QCD evolution brings these numbers down, similarly to the case of SQM.

For the tensor form factor we find
\begin{align}
f_{\perp\gamma}^{t}\left(  q^{2}\right)   &  =\frac{(M^{2}+\Lambda^{2})\left[
2(2\Lambda^{2}+S_{0})\mathrm{tanh}^{-1}\frac{q}{\sqrt{S}}-2\sqrt{S_{0}%
S}\mathrm{tanh}^{-1}\frac{q}{S_{0}}+q\sqrt{S}\log\left(  1+\frac{\Lambda^{2}%
}{M^{2}}\right)  \right]  }{q\sqrt{S}\left[  (M^{2}+\Lambda^{2})\log\left(
1+\frac{\Lambda^{2}}{M^{2}}\right)  -\Lambda^{2}\right]  },\\
S  &  =4M^{2}+4\Lambda^{2}+q^{2},\;\;S_{0}=4M^{2}+4\Lambda^{2},
\end{align}
which satisfies (\ref{48}). Other form factors involve numerical integration.
All form factors obtained in the NJL model are qualitatively very similar to
the case of SQM shown in Fig.~\ref{Normwb}.

\section{Photon light-cone wave functions in quark models}

We use the light-cone coordinates with the convention $a_{\pm}=a_{0}\pm a_{3}%
$. The null vector $n^{\mu}=(n^{0},n^{1},n^{2},n^{3})=(1/p_{+},0,0,-1/p_{+})$
satisfies the conditions $n^{2}=0$ and $n\cdot p=1$, $n\cdot a=a_{+}/p_{+}$.
We also denote
\[
\frac{d^{4}k}{(2\pi)^{4}}=\frac{dk_{+}dk_{-}d^{2}k_{\perp}}{2(2\pi)^{4}}\equiv
d\tilde{k}.
\]
The definition (\ref{phi}) is equivalent to (from now on we only keep the
leading twist contribution)
\begin{equation}
ie_{q}\chi_{\mathrm{m}}\langle\bar{q}q\rangle f_{\perp\gamma}^{t}\left(
q^{2}\right)  \left(  q_{\beta}e_{\alpha}^{(\lambda)}-q_{\alpha}e_{\beta
}^{(\lambda)}\right)  \phi_{\perp\gamma}(x)=\int_{-\infty}^{\infty}\frac
{d\tau}{\pi}e^{-i\tau(2x-1)}\langle0|\bar{q}(\tau n)\sigma_{\alpha\beta
}q(-\tau n)|\gamma^{\lambda}(q)\rangle. \label{v1}%
\end{equation}
In the one loop approximation the quark model evaluation of the matrix element
of the quark bilinear yields
\begin{equation}
\chi_{\mathrm{m}}\langle\bar{q}q\rangle f_{\perp\gamma}^{t}\left(
q^{2}\right)  \left(  q_{\beta}e_{\alpha}^{(\lambda)}-q_{\alpha}e_{\beta
}^{(\lambda)}\right)  \phi_{\perp\gamma}(x)=-iN_{c}\int d\tilde{k}%
\delta(k\cdot n-x)\mathrm{Tr}[\sigma_{\alpha\beta}S(k)\Gamma\cdot
e^{(\lambda)}S(k-q)], \label{phDA}%
\end{equation}
where the trace is over the Dirac space. The interpretation of this result is
clear: the photon DA times a Lorentz structure is equal to the quark-loop
integral, where the $+$-component of the momentum of one of the quarks is
constrained to $k_{+}=xp_{+}$ (see Fig. \ref{diag}). If one does not carry out
the transverse momentum integral in Eq.~(\ref{phDA}), then one obtains the
photon \emph{light-cone wave function}, $\Phi_{\perp \gamma}(x,k_{\perp})$,
\begin{equation}
\chi_{\mathrm{m}}\langle\bar{q}q\rangle f_{\perp\gamma}^{t}\left(
q^{2}\right)  \left(  q_{\beta}e_{\alpha}^{(\lambda)}-q_{\alpha}e_{\beta
}^{(\lambda)}\right)  \Phi_{\perp\gamma}(x,k_{\perp})=-iN_{c}\int\frac
{dk_{+}dk_{-}}{2(2\pi)^{4}}\delta(k\cdot n-x)\mathrm{Tr}[\sigma_{\alpha\beta
}S(k)\Gamma\cdot e^{(\lambda)}S(k-q)]. \label{phlcwf}%
\end{equation}
Obviously, we have
\[
\int d^{2}k_{\perp}\Phi_{\perp\gamma}(x,k_{\perp})=\phi_{\perp\gamma}(x).
\]
For the twist-3 vector component of the photon light-cone wave function and DA
we find from the definition (\ref{phiV}), with analogous steps as above, the
expression
\begin{equation}
e_{\perp\mu}^{(\lambda)}\Psi_{\gamma\parallel}^{(v)}(x,k_{\perp})+q_{\mu
}n\cdot e^{(\lambda)}\Psi_{\gamma\perp}^{(v)}(x,k_{\perp})=\frac{N_{c}%
}{f_{3\gamma}}\int\frac{dk_{+}dk_{-}}{2(2\pi)^{4}}\delta(n\cdot
k-x)\mathrm{Tr}[\gamma_{\mu}S(k)\Gamma\cdot e^{(\lambda)}S(k-q)],
\label{phlcwfV}%
\end{equation}
while the DAs are $\psi_{\gamma i}^{(v)}(x)=\int d^{2}k_{\perp}\Psi_{\gamma
i}^{(v)}(x,k_{\perp})$, with $i=(\parallel,\perp)$.

The case of the axial twist-3 component is somewhat more complicated, since
the definition (\ref{phiA}) involves a power of the coordinate $z$ in the
tensor structure. As a result we find
\begin{equation}
e_{q}f_{3\gamma}f_{\gamma}^{a}\left(  q^{2}\right)  \epsilon
^{\mu\nu\alpha\beta}e_{\nu}^{(\lambda)}q_{\alpha}n_{\beta}\psi_{\gamma}%
^{(a)}(x)=\hspace{-2mm}\int_{-\infty}^{\infty}\frac{d\tau}{\pi(\tau
+i\epsilon)}e^{-i\tau(2x-1)}\langle0|\overline{q}(\tau n)\gamma^{\mu}%
\gamma_{5}q(-\tau n)|\gamma^{\lambda}(q)\rangle. \label{phDAax}%
\end{equation}
The one-quark loop evaluation yields
\begin{equation}
e_{q}f_{3\gamma}f_{\gamma}^{a}\left(  q^{2}\right)  \epsilon
^{\mu\nu\alpha\beta}e_{\nu}^{(\lambda)}q_{\alpha}n_{\beta}\psi_{\gamma}%
^{(a)}(x)=\frac{4N_{c}}{f_{3\gamma}}\int d\tilde{k}\Theta(n\cdot
k-x)\mathrm{Tr}[\gamma^{\mu}\gamma_{5}S(k)\Gamma\cdot e^{(\lambda)}S(k-q)].
\end{equation}
We may introduce the derivative of the distribution, $d/dx\,\psi_{\gamma
}^{(a)}(x)$, for which we have
\begin{equation}
e_{q}f_{3\gamma}f_{\gamma}^{a}\left(  q^{2}\right)  \epsilon
^{\mu\nu\alpha\beta}e_{\nu}^{(\lambda)}q_{\alpha}n_{\beta}\frac{d}{dx}%
\psi_{\gamma}^{(a)}(x)=\frac{4N_{c}}{f_{3\gamma}}\int d\tilde{k}\delta(n\cdot
k-x)\mathrm{Tr}[\gamma^{\mu}\gamma_{5}S(k)\Gamma\cdot e^{(\lambda)}S(k-q)].
\label{phDAaxd}%
\end{equation}
The evaluation of this quantity is similar to the evaluation of $\phi_{\gamma
}$ or $\psi^{(v)}$. The axial light-cone wave function is obtained from
Eq.~(\ref{phDAaxd}) if the $k_{\perp}$ integration is left out, in analogy to
Eqs.~(\ref{phlcwf},\ref{phlcwfV}).

\section{Photon light-cone wave functions in chiral quark models}

\subsection{Leading-twist photon DAs}

\begin{figure}[tb]
\begin{center}
\includegraphics[width=9cm]{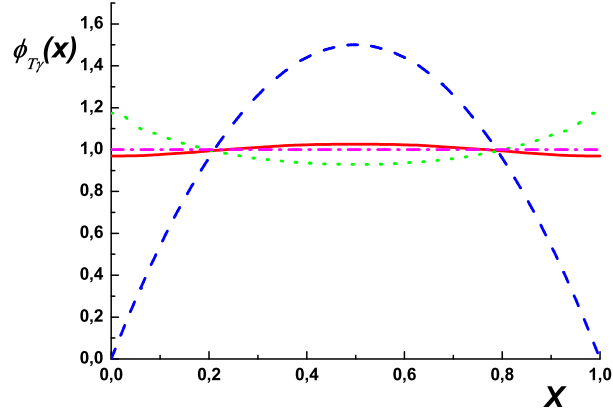}
\end{center}
\caption{The leading-twist photon distribution amplitude in the tensor
channel, $\phi_{\perp\gamma}(x,q^{2}=0)$, evaluated at the quark model scale
in the non-local quark
model -- solid line, in the spectral quark model and the NJL model
($=1$) -- dot dashed line, its asymptotic form -- dashed line, and the result
of the local approximation to the instanton model -- dotted line.}%
\label{F2V}%
\end{figure}

The distribution amplitudes of the real photon calculated in the instanton
model in the chiral limit may be cast in a closed form. It is convenient to
introduce notations for the integration variables\footnote{Some details of
calculation method first developed in \cite{Dorokhov:1998up,Dorokhov:2000gu}
and the relevant expressions for the integrals are given in Appendix
\ref{app:form} and \ref{app:Gauss}.}%
\begin{equation}
u_{+}=u-i\lambda x,\quad u_{-}=u+i\lambda\overline{x},\quad M_{\pm}=M\left(
u_{\pm}\right)  ,\quad D_{\pm}=D\left(  u_{\pm}\right)  ,\quad\overline
{x}=1-x.
\end{equation}
By using the definitions of Sect.~IV one gets the twist-2 DA for the
$\sigma_{\mu\nu}$ structure (Fig. \ref{F2V})
\begin{equation}
\phi_{\perp\gamma}\left(  x,q^{2}=0\right)  =\frac{1}{\chi_{\mathrm{m}}}%
\frac{N_{c}}{4\pi^{2}}\left[  \Theta(\overline{x}x)\int_{0}^{\infty}%
du\frac{M\left(  u\right)  }{D\left(  u\right)  }-\int_{0}^{\infty}%
du\int_{-\infty}^{\infty}\frac{d\lambda}{2\pi}\frac{M_{+}M_{-}}{D_{+}D_{-}%
}M^{\left(  1\right)  }\left(  u_{+},u_{-}\right)  \right]  . \label{PhiTt}%
\end{equation}

Within the non-local model we can find from (\ref{PhiTt}) the following
light-cone wave function:
\begin{equation}
\Phi_{\perp\gamma}\left(  x,\mathbf{k}_{\perp}^{2},q^{2}=0\right)  =\frac
{1}{\chi_{\mathrm{m}}}\frac{N_{c}}{4\pi^{2}}\left[  \Theta(\overline{x}%
x)\frac{M\left(  \text{\textbf{k}}_{\perp}^{2}\right)  }{D\left(
\mathbf{k}_{\perp}^{2}\right)  }-\int_{-\infty}^{\infty}\frac{d\lambda}{2\pi
}\frac{M_{+}M_{-}}{D_{+}D_{-}}M^{\left(  1\right)  }\left(  u_{+}%
,u_{-}\right)  \right]  .
\end{equation}
It exhibits an exponential decay law at large $\mathbf{k}_{\perp}^{2}$, which
is a consequence of the nonlocality of the model \cite{Dorokhov:2005pg} and is
in accordance with the conclusion made in \cite{Teryaev:2004df} that
finiteness of all $\mathbf{k}_{\perp}$ moments of the quark distributions
guarantees the exponential fall-off of the cross sections.

In the vector channel the leading twist structure for the real photon
$\phi_{\parallel\gamma}\left(  x,q^{2}=0\right)  $ is decoupled but the DA is
not zero itself and has form
\begin{equation}
\phi_{\parallel \gamma}\left(  x,q^{2}=0\right)  =\Theta(\overline{x}x)
\label{wb1}%
\end{equation}
independently on the nonlocality shape.

The results for SQM and NJL models are obtained from (\ref{PhiTt}) by taking
the local limit (all derivatives are zero $M^{\prime}=0$, $M^{(1)}=0$,
\emph{etc.} ). This gives the result
\begin{equation}
\phi_{i\gamma}\left(  x,q^{2}=0\right)  =\Theta(\overline{x}x), \label{PhiTsm}%
\end{equation}
for $i=\perp,\parallel$. These leading twist DAs calculated within the
instanton and local models are shown in Fig. \ref{F2V} along with its
asymptotic form defined at very large scale $\mu$%
\begin{equation}
\phi_{\perp\gamma}^{\mathrm{asympt}}\left(  x\right)  =6x\overline{x}.
\label{F2Vasympt}%
\end{equation}

In Fig. \ref{F2V} there is also given the result obtained in
\cite{Petrov:1998kg} in gauge noninvariant approximation
\begin{align}
\phi_{\perp\gamma}^{\mathrm{approx}}\left(  x\right)   &  =\frac{1}%
{\chi_{\mathrm{m}}^{^{\prime}}}\frac{N_{c}}{4\pi^{2}}\int_{0}^{\infty}%
du\int_{-\infty}^{\infty}\frac{d\lambda}{2\pi}\frac{\left(  1-x\right)
M_{+}+xM_{-}}{D_{+}D_{-}},\label{F2Vddp}\\
\chi_{\mathrm{m}}^{\mathrm{approx}}  &  =\frac{N_{c}}{4\pi^{2}}\int
du\frac{u\left(  M\left(  u\right)  -\frac{1}{2}uM^{\prime}\left(  u\right)
\right)  }{D^{2}\left(  u\right)  }\nonumber
\end{align}
when the full vector vertex (\ref{vert1}) is substituted by the local part.
This approximation resulted in a violation of the proper normalization as
implied by the Ward-Takahashi identities.

\begin{figure}[th]
\hspace*{0.1cm} \begin{minipage}{8.5cm}
\vspace*{0.5cm} \epsfxsize=6cm \epsfysize=5cm \centerline{\epsfbox
{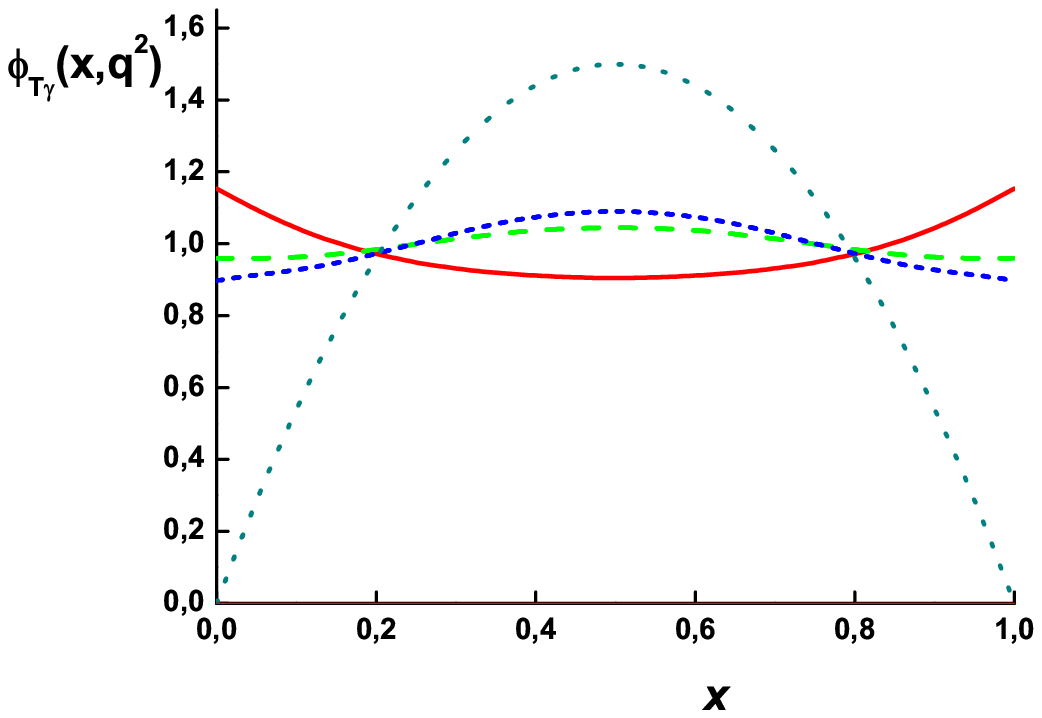}}
\caption[dummy0]{Dependence of the twist-2 tensor component of the photon DA on transverse
momentum squared ($q^{2}=0.25$ GeV$^{2}$ solid line, $q^{2}=0$ GeV$^{2}$
dashed line, $q^{2}=-0.09$ GeV$^{2}$ short-dashed line, asymptotic DA - dotted line). Nonlocal model at the
quark model scale.}
\label{TensorQ}
\end{minipage}\hspace*{0.5cm} \begin{minipage}{8.5cm}
\vspace*{0.5cm} \epsfxsize=6cm \epsfysize=5cm \centerline{\epsfbox
{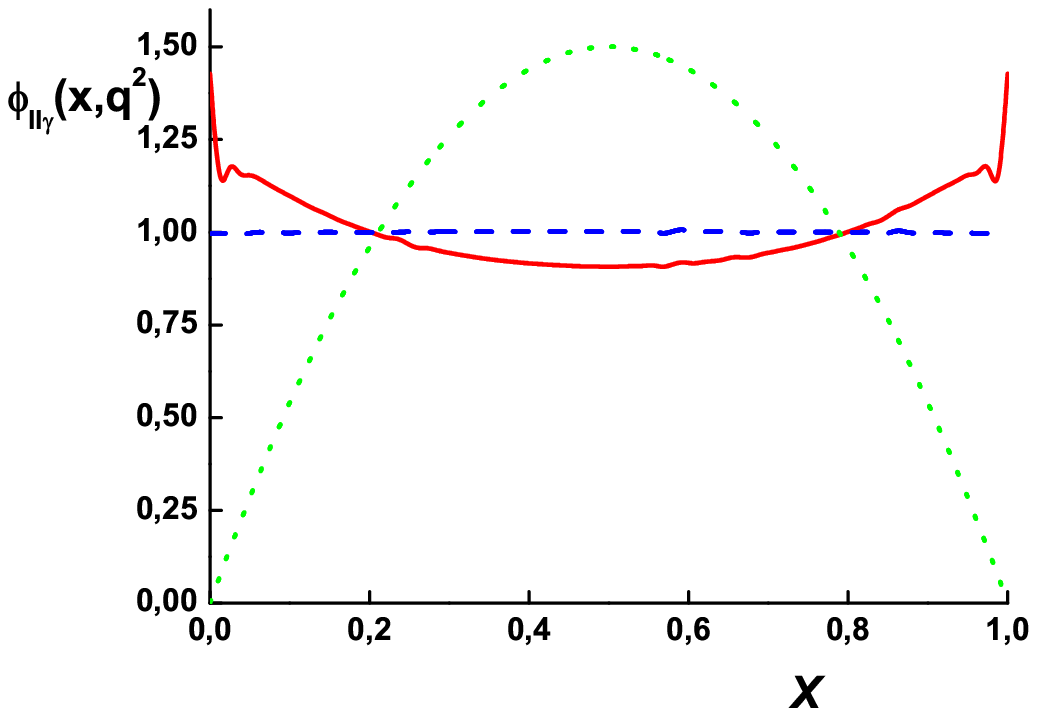}}
\caption[dummy0]{Same as Fig. \ref{TensorQ} for the twist-2 vector component of the photon DA.}
\label{FVii}
\end{minipage}
\end{figure}

The dependence of the leading twist photon DA $\phi_{\perp\gamma}\left(
x,q^{2}\right)  $ on the photon virtuality in space-like and time-like
regions is demonstrated in Fig. \ref{TensorQ}. Both leading twist DAs are
normalized by unity%
\begin{equation}
\int_{0}^{1}\phi_{\gamma}\left(  x,q^{2}\right)  =1.
\end{equation}
At very high space-like photon virtuality the photon DA approaches the $\delta$-function
distribution concentrated
at the edge points of the $x$ interval. It corresponds to the configuaration when all momenta pass
through quark or anti-quark line, with other line being soft. At time-like virtualities
approaching the $\rho$-meson pole
the photon DA will be concentrated in the middle of $x$ interval.

\subsection{Spectral quark model}

The expressions for SQM are obtained from the formulas of the preceding
sections with the formal replacement $M(p)\rightarrow\omega$, and supplying
the loop integrals with the addition spectral integration $\int_{C}\rho
(\omega)d\omega$, as explained in Sect.~III.B. As the result, we find the
following expression for the leading-twist component of the light-cone wave
function for the real photon in the tensor channel:
\begin{equation}
\Phi_{\perp\gamma}%
(x,\mathchoice{\mbox{\boldmath$\displaystyle k$}} {\mbox{\boldmath$\textstyle k$}} {\mbox{\boldmath$\scriptstyle k$}} {\mbox{\boldmath$\scriptscriptstyle k$}}_{\perp
})=\frac{6}{m_{\rho}^{2}%
(1+4\mathchoice{\mbox{\boldmath$\displaystyle k$}} {\mbox{\boldmath$\textstyle
k$}} {\mbox{\boldmath$\scriptstyle k$}} {\mbox{\boldmath$\scriptscriptstyle
k$}}_{\perp}^{2}/m_{\rho}^{2})^{5/2}},
\end{equation}
Note the power-law fall-off at large transverse momenta, $\Phi_{\gamma
}%
(x,\mathchoice{\mbox{\boldmath$\displaystyle k$}}{\mbox{\boldmath$\textstyle
k$}} {\mbox{\boldmath$\scriptstyle k$}}{\mbox{\boldmath$\scriptscriptstyle
k$}}_{\perp})\sim1/k_{\perp}^{5}$. In cross section this leads to tails of the
form $\sim1/k_{\perp}^{10}$.

For the virtual photon of virtuality $q^{2}$, the tensor component of the
light-cone wave function has the form
\begin{equation}
\Phi_{\perp\gamma^{\ast}}(x,\mathchoice{\mbox{\boldmath$\displaystyle
k$}} {\mbox{\boldmath$\textstyle k$}} {\mbox{\boldmath$\scriptstyle k$}} {\mbox{\boldmath$\scriptscriptstyle k$}}_{\perp
})=\frac{6\left(  1+\frac{q^{2}}{m_{\rho}^{2}}\right)  }{m_{\rho}^{2}\left(
1+4\frac
{\mathchoice{\mbox{\boldmath$\displaystyle k$}} {\mbox{\boldmath$\textstyle
k$}} {\mbox{\boldmath$\scriptstyle k$}} {\mbox{\boldmath$\scriptscriptstyle
k$}}_{\perp}^{2}+q^{2}x(1-x)}{m_{\rho}^{2}}\right)  ^{5/2}},
\end{equation}
We also find
\begin{equation}
\int_{0}^{1}dx\Phi_{\perp\gamma^{\ast}}%
(x,\mathchoice{\mbox{\boldmath$\displaystyle k$}} {\mbox{\boldmath$\textstyle k$}} {\mbox{\boldmath$\scriptstyle k$}} {\mbox{\boldmath$\scriptscriptstyle k$}}_{\perp
})=\frac{2m_{\rho}(m_{\rho}^{2}+q^{2})(3m_{\rho}^{2}+q^{2}%
+12\mathchoice{\mbox{\boldmath$\displaystyle k$}} {\mbox{\boldmath$\textstyle
k$}} {\mbox{\boldmath$\scriptstyle k$}} {\mbox{\boldmath$\scriptscriptstyle
k$}}_{\perp}^{2})}{(m_{\rho}^{2}+4\mathchoice{\mbox{\boldmath$\displaystyle
k$}} {\mbox{\boldmath$\textstyle k$}} {\mbox{\boldmath$\scriptstyle k$}} {\mbox{\boldmath$\scriptscriptstyle k$}}_{\perp
}^{2})^{3/2}(m_{\rho}^{2}+q^{2}+4\mathchoice{\mbox{\boldmath$\displaystyle
k$}} {\mbox{\boldmath$\textstyle k$}} {\mbox{\boldmath$\scriptstyle k$}} {\mbox{\boldmath$\scriptscriptstyle k$}}_{\perp
}^{2})^{2}}.
\end{equation}
The transverse-momentum rms is
\begin{equation}
\langle
\mathchoice{\mbox{\boldmath$\displaystyle k$}} {\mbox{\boldmath$\textstyle k$}} {\mbox{\boldmath$\scriptstyle k$}} {\mbox{\boldmath$\scriptscriptstyle k$}}_{\perp
}^{2}\rangle\equiv\frac{\int d^{2}k_{\perp}%
\mathchoice{\mbox{\boldmath$\displaystyle k$}} {\mbox{\boldmath$\textstyle k$}} {\mbox{\boldmath$\scriptstyle k$}} {\mbox{\boldmath$\scriptscriptstyle k$}}_{\perp
}^{2}\Phi_{\perp\gamma^{\ast}}%
(x,\mathchoice{\mbox{\boldmath$\displaystyle k$}} {\mbox{\boldmath$\textstyle
k$}} {\mbox{\boldmath$\scriptstyle k$}} {\mbox{\boldmath$\scriptscriptstyle
k$}}_{\perp})}{\phi_{\perp\gamma^{\ast}}(x)}=\frac{m_{\rho}^{2}}{2}%
+2q^{2}x(1-x).
\end{equation}
For the real photon one has the estimate $\langle
\mathchoice{\mbox{\boldmath$\displaystyle k$}} {\mbox{\boldmath$\textstyle
k$}}{\mbox{\boldmath$\scriptstyle k$}}{\mbox{\boldmath$\scriptscriptstyle
k$}}_{\perp}^{2}\rangle=(544$ MeV$)^{2}$. For the instanton model one
gets
\begin{equation}
\langle
\mathchoice{\mbox{\boldmath$\displaystyle k$}} {\mbox{\boldmath$\textstyle k$}} {\mbox{\boldmath$\scriptstyle k$}} {\mbox{\boldmath$\scriptscriptstyle k$}}_{\perp
}^{2}\rangle\equiv
\left(  \int du\frac{M\left(  u\right)  }{D\left(  u\right)  }\right)
^{-1}
\int du\frac{uM\left(  u\right)  }{D\left(  u\right)
}=(600~\mathrm{MeV})^{2}.
\end{equation}
For the real photon DA one finds the constant form (\ref{PhiTsm}),
while for the virtual case we find
the following normalized DA:
\[
\phi_{\perp\gamma^{\ast}}(x)\equiv\int d^{2}k_{\perp}\Phi_{\perp\gamma
}%
(x,\mathchoice{\mbox{\boldmath$\displaystyle k$}} {\mbox{\boldmath$\textstyle k$}} {\mbox{\boldmath$\scriptstyle k$}} {\mbox{\boldmath$\scriptscriptstyle k$}}_{\perp
})=\frac{\left(  1+\frac{q^{2}}{m_{\rho}^{2}}\right)  }{\left(  1+4\frac
{q^{2}x(1-x)}{m_{\rho}^{2}}\right)  ^{3/2}}.
\]

The corresponding expressions for the $\rho$ meson are obtained by taking the limit
$q^{2}\rightarrow-m_{\rho}^{2}$, which for the DA yields
\begin{equation}
\phi_{\perp\rho}(x)=\delta(x-\frac{1}{2}) \label{del12}%
\end{equation}
This highly singular behavior is washed out by the QCD evolution, see
Sect.~\ref{evolve}.

The leading twist photon amplitude in the vector channel, $\phi_{\parallel \gamma}(x)$, is of the form given
in Eq.~(\ref{PhiTsm}), \emph{i.e.} constant.

\subsection{NJL model}

In the NJL model one obtains the same leading-twist DAs for the real photon as
Eq.~(\ref{PhiTsm}), \emph{i.e.} $\phi_{i\gamma}\left(  x,q^{2}=0\right)
=\Theta(\overline{x}x)$. Extension to the $\rho$-meson pole, however, is
problematic due to appearance of the quark production threshold. For this
reason we do not perform this analysis. Note that SQM is free of this problem,
as no quark poles are present in this model -- a feature sometimes referred to
as \emph{analytic confinement}.

\subsection{QCD evolution}

\label{evolve} \begin{figure}[b]
\begin{center}
\subfigure{\includegraphics[angle=0,width=0.4\textwidth]{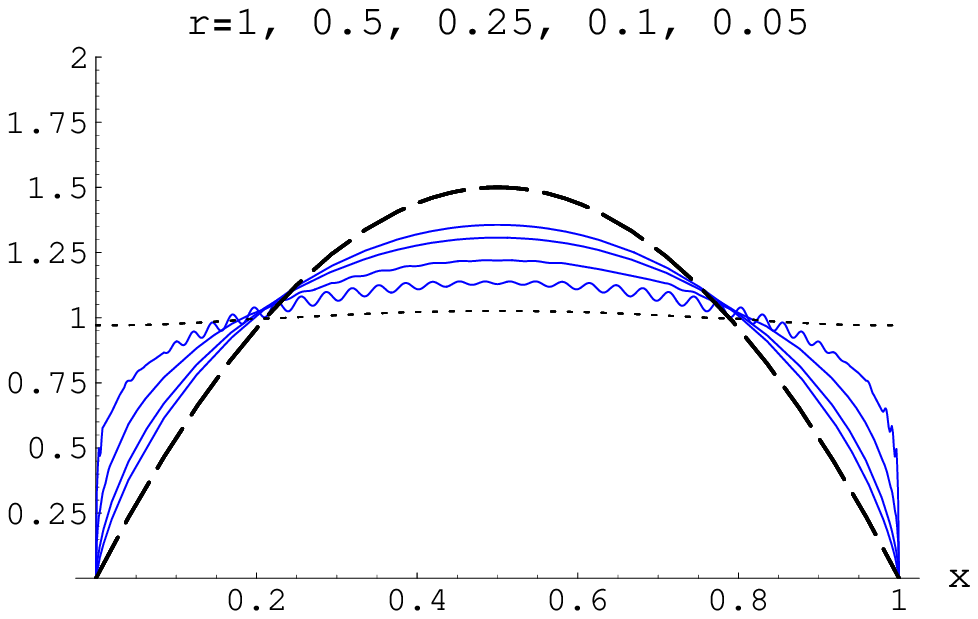}}
\subfigure{\includegraphics[angle=0,width=0.4\textwidth]{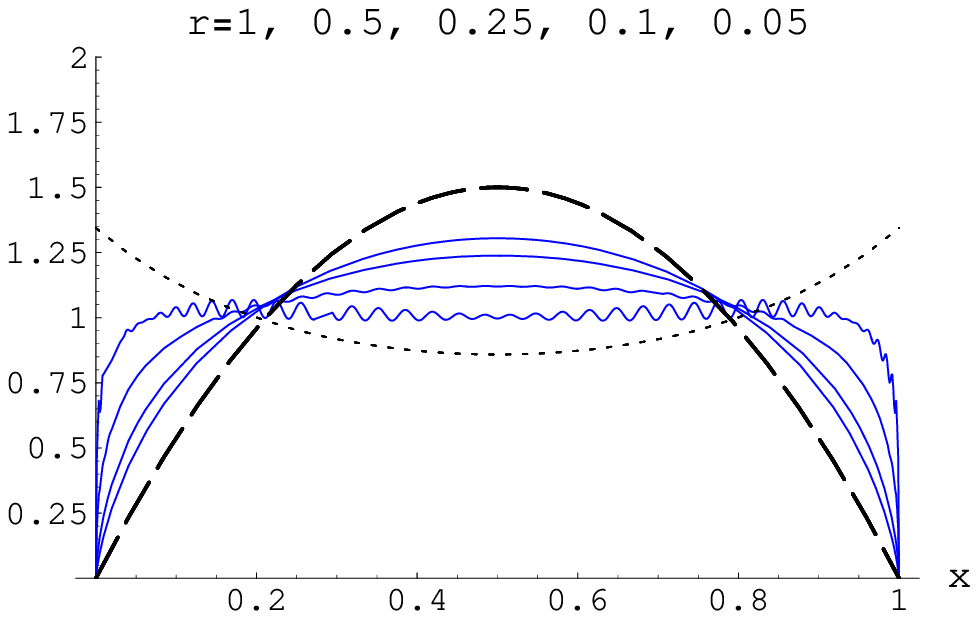}}\\[0pt]%
\hspace{-4mm}
\subfigure{\includegraphics[angle=0,width=0.4\textwidth]{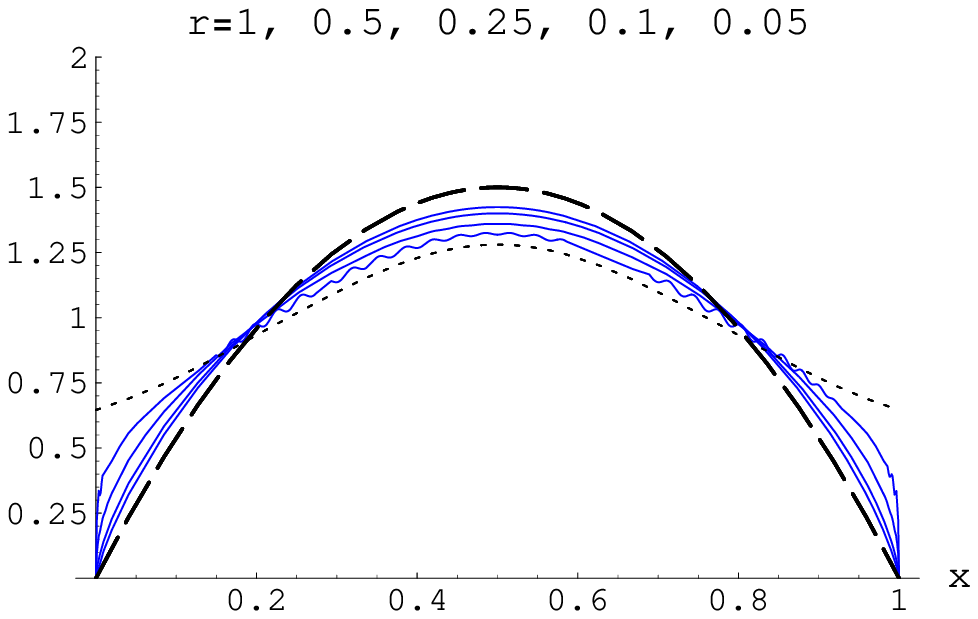}}
\hspace{-3mm}
\end{center}
\caption{The LO ERBL evolution of the nonlocal model predictions for the
leading-twist \emph{tensor} photon DA $\phi_{\perp\gamma}^{(t)}(x,q^{2})$. Top
left: the real photon DA, top right: the virtual photon at $q^{2}=0.25$
GeV$^{2}$; bottom: the virtual photon at $q^{2}=-0.09$ GeV$^{2}$. The dashed
lines show the asymptotic DA, $6x(1-x)$. Initial conditions, indicated by
dotted lines, are evaluated in the nonlocal quark model at the initial scale
$\mu^{\mathrm{inst}}=530$~MeV. The solid lines correspond to evolved DA'a at
scales $Q=1$, $2.4$, $10$, and $1000$~GeV. The corresponding values of the evolution ratio $r$
are given in the figures. The appearance of tiny wiggles is a numerical
artefact. }%
\label{NLQMevT}%
\end{figure}\begin{figure}[bb]
\begin{center}
\subfigure{\includegraphics[angle=0,width=0.45\textwidth]{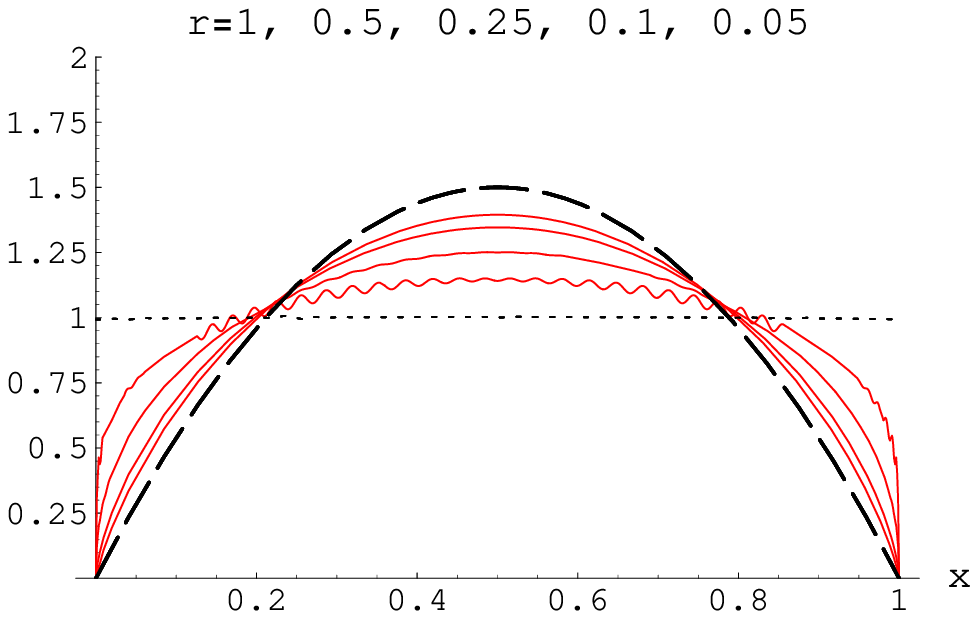}}
\subfigure{\includegraphics[angle=0,width=0.45\textwidth]{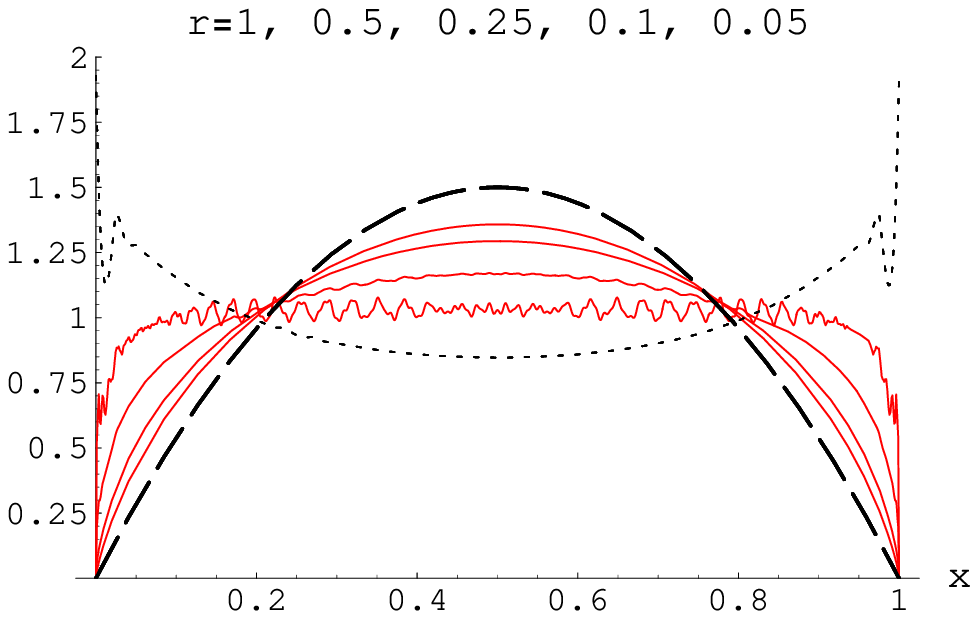}}\\[0pt]%
\hspace{-4mm}
\end{center}
\caption{The LO ERBL evolution of the nonlocal model predictions for the
leading-twist \emph{vector} DAs of the photon $\phi_{\parallel\gamma}%
^{(v)}(x,q^{2})$. Left: the real photon DA; right: the virtual photon at
$q^{2}=0.25$ GeV$^{2}$. The dashed lines show the asymptotic DA, $6x(1-x)$.
Initial conditions, indicated by dotted lines, are evaluated in the nonlocal
quark model at the initial scale $\mu^{\mathrm{inst}}=530$~MeV. The solid
lines correspond to evolved DA'a at scales $Q=1$, $2.4$, $10$, and $1000$~GeV.
}%
\label{NLQMevV}%
\end{figure}\begin{figure}[bbb]
\begin{center}
\subfigure{\includegraphics[angle=0,width=0.455\textwidth]{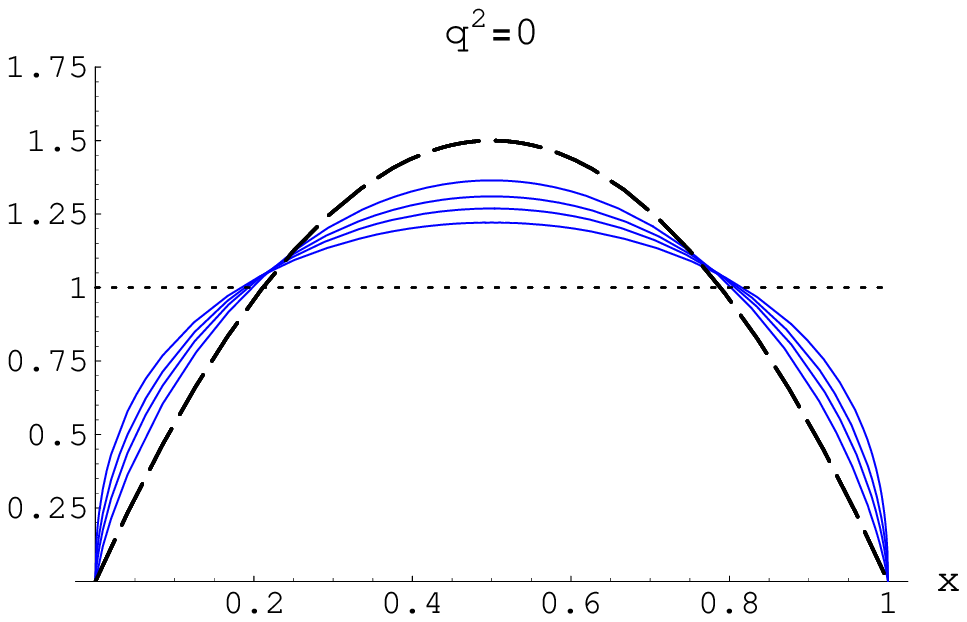}}
\subfigure{\includegraphics[angle=0,width=0.475\textwidth]{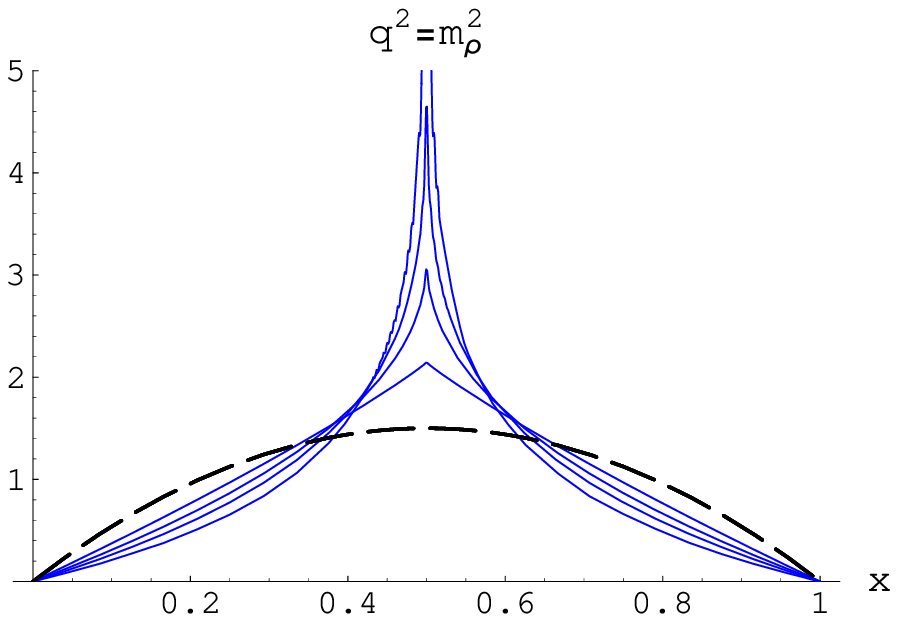}}\\[0pt]%
\hspace{-4.0mm}
\subfigure{\includegraphics[angle=0,width=0.455\textwidth]{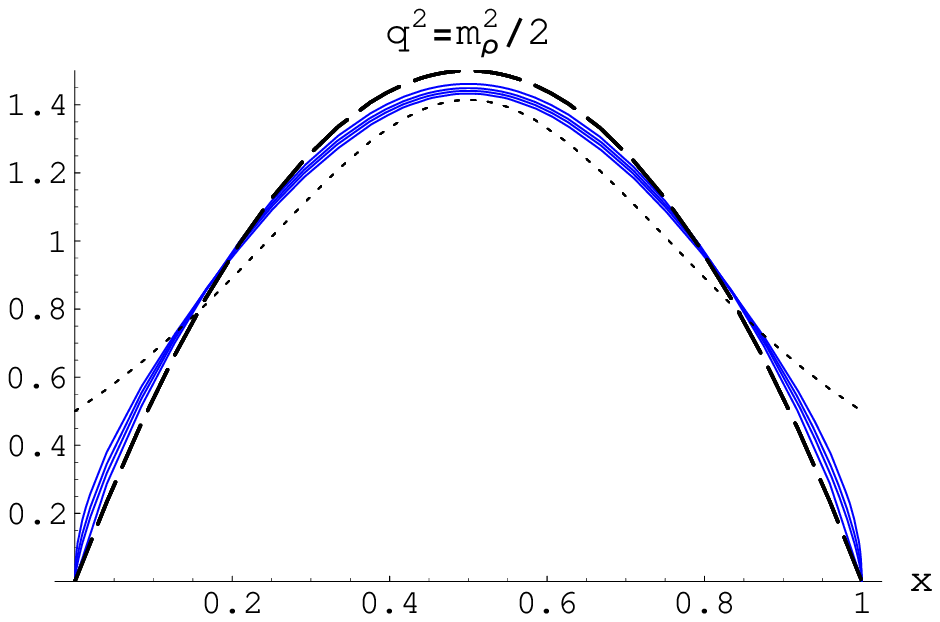}}
\hspace{-3.0mm}
\subfigure{\includegraphics[angle=0,width=0.455\textwidth]{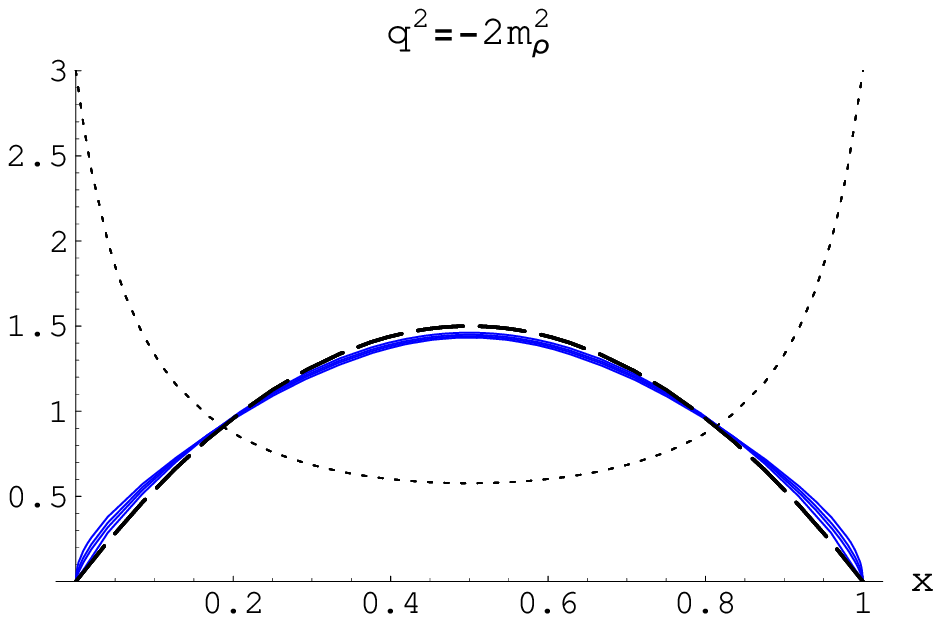}}
\end{center}
\caption{Evolution of the real photon DA (top left), $\rho$-meson DA (top
right), virtual photon at $q^{2}=-m_{\rho}^{2}/2$ (bottom left), and virtual
photon at $q^{2}=2m_{\rho}^{2}$ (bottom right). Initial conditions, indicated
by dotted lines, are evaluated in the spectral quark model at the initial
scale $\mu^{\mathrm{SQM}}=313$~MeV. For the $\rho$-meson case the initial
condition is given in Eq.~(\ref{del12}). The solid lines correspond to evolved
DA'a at scales $Q=1$, $2.4$, $10$, and $1000$~GeV. With the larger the scale
the evolved DA is closer to the asymptotic form $6x(1-x)$, plotted with the
dashed line. The corresponding values of the evolution ratio $r$
are given in the figures. The appearance of tiny wiggles is a numerical
artefact. }%
\label{fig:SQMev}%
\end{figure}

The results of the previous sections referred to the low-energy scale of the
quark model, $\mu$. In order to relate to results at higher scales, the QCD
evolution is necessary. Within the effective model approach the method has
been described in detail in Ref.~\cite{RuizArriola:2002bp} and is similar to
the case of parton distribution evolution
\cite{Davidson:1994uv,Dorokhov:1998up}. For the twist-2 photon and $\rho
$-meson DAs the leading-order QCD evolution is made on the basis of the
Gegenbauer polynomials. The basic logic here is that the quark models provide
the initial condition for the QCD evolution, now written as $\phi^{i}%
(x,\mu^{2})$. The evolved leading-twist distribution amplitudes
read~\cite{Mueller:1994cn}
\begin{equation}
\phi^{i}(x,q^{2})=\phi_{\mathrm{as}}^{i}(x)\sum{^{\prime}}_{n=0}^{\infty}%
C_{n}^{\lambda}(2x-1)a_{n}^{i}(q^{2}),
\end{equation}
with
\[
\phi_{\mathrm{as}}^{\mathrm{T}}(x)=\phi_{\mathrm{as}}^{\mathrm{V}}%
(x)=\phi_{\mathrm{as}}^{\mathrm{AV}}(x)=6x\bar{x},
\]
$C_{n}^{\lambda}$ denoting the Gegenbauer polynomials, the prime indicating
that only even $n$ enter the sum, and $\lambda=3/2$ for T, V, and
AV. The Gegenbauer coefficients $a_{n}^{i}$ evolve with the scale. The
corresponding formulas for the leading-order QCD evolution are given in
App.~\ref{app:QCD}.
We note that the exponent in Eq.~(\ref{rat}) contains the difference
$(\gamma_{n}^{i}-\gamma_{0}^{i})$, such as to make the zeroth moment (the norm) of
the DAs constant.

Our methods of determining the {\it a priori} unknown quark-model scale $\mu$ have
been presented in \cite{Dorokhov:2005pg} for the case of the nonlocal model and
in Ref.~\cite{RuizArriola:2002bp} for the local models.

The numerical analysis of the evolution of the tensor DA for the photon and
the $\rho$-meson proceeds exactly as written in the above formulas, with the
series approximated by finite sums over $n$. Numerically, about 100 terms are
needed to achieve the accuracy in the presented figures. The results for the
twist-2 photon DAs are shown in Figs.~\ref{NLQMevT} and \ref{NLQMevV}.

The results for SQM are shown in Fig.~\ref{fig:SQMev}. We show the evolution
of the DAs for the real photon, for the $\rho$ meson, as well as for virtual
photon. We note that the evolution radically changes the character of the
curves. For the real photon one starts with the flat DA of Eq.~(\ref{PhiTsm}),
exactly as for the case of the pion in the local model
\cite{RuizArriola:2002bp}, with the evolution providing vanishing of the
amplitude at the end-points and gradually approaching the asymptotic limit. So
it means that all moments of the DAs evolve as continuous function, but the DA
itself at some points (edge points in the discussed case) evolve
discontinuously. For the $\rho$-meson case the result is even more dramatic,
with the initial condition having the singular form of Eq.~(\ref{del12}),
evolving to broader regular functions and also tending at large evolution
scales $Q$ to the asymptotic limit. We note that this evolution is very slow.
The lower part of Fig.~\ref{fig:SQMev} shows the DA for the virtual photon
with two selected virtualities: time-like $q^{2}=-m_{\rho}^{2}/2$ and
space-like $q^{2}=2m_{\rho}^{2}$. We have to note that because of low initial
evolution scale the QCD\ evolution carried out at two-loop level would give
more realistic results.

The results for the real photon in the NJL model are the same as in SQM.

\section{Higher twist components}

\subsection{Non-leading twist photon DAs in the non-local chiral quark model}

\begin{figure}[h]
\hspace*{0.1cm} \begin{minipage}{9cm}
\vspace*{0.5cm} \epsfxsize=6cm \epsfysize=5cm \centerline{\epsfbox
{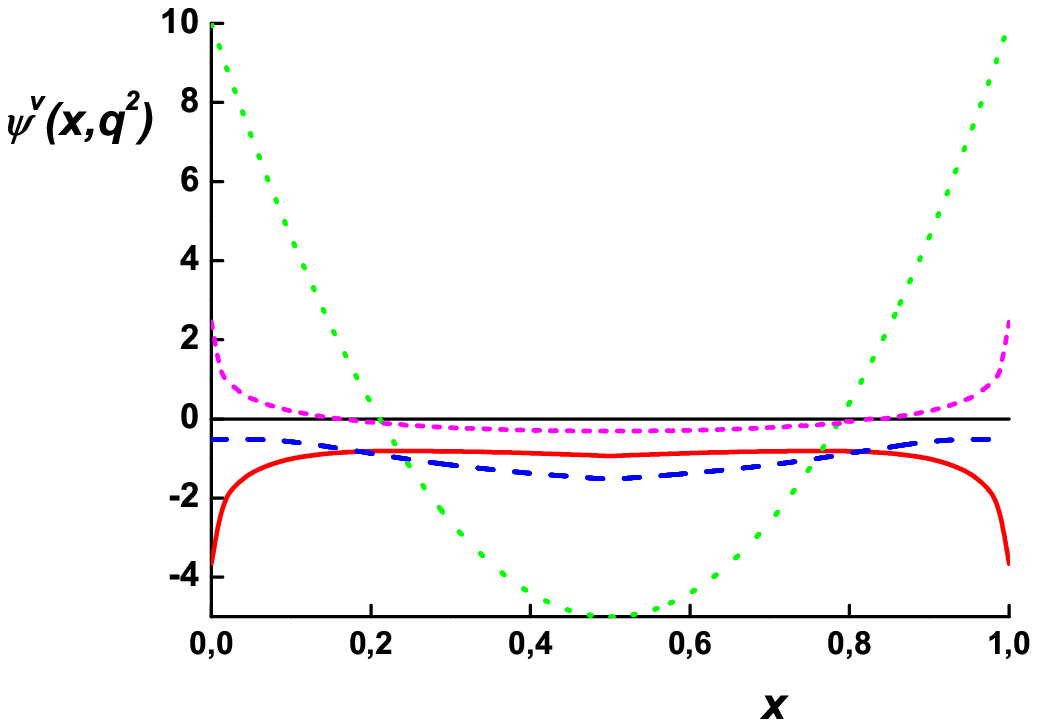}}
\caption[dummy0]{Dependence of the regular part (without $\delta$-functions) of the twist-3 vector component of the photon DA on
transverse momentum squared ($q^2=0.25$ GeV$^2$ solid line, $q^2=0$ GeV$^2$ dashed line).
The short dashed line corresponds to the WW approximation (\ref{WWv}) and
the dotted line is for the conformal approximation (\ref{PsiVconf}) }
\label{V3}
\end{minipage}\hspace*{0.5cm} \begin{minipage}{9cm}
\vspace*{0.5cm} \epsfxsize=6cm \epsfysize=5cm \centerline{\epsfbox
{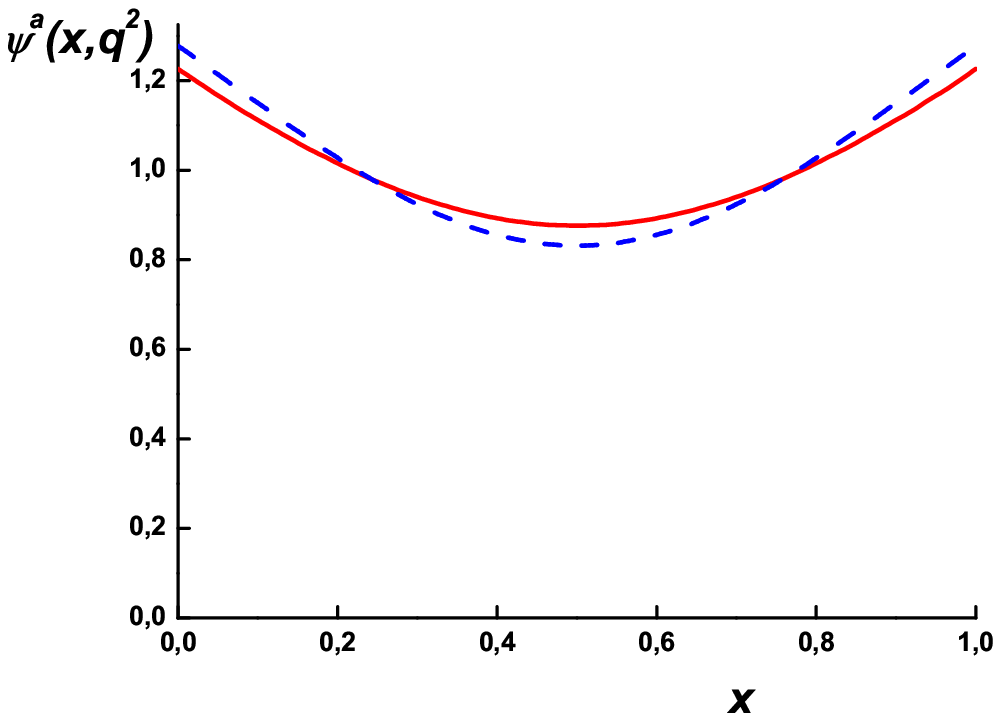}}
\caption[dummy0]{Dependence of the twist-3 axial component of the photon DA on
transverse momentum squared ($q^2=0.25$ GeV$^2$ solid line, $q^2=0$ GeV$^2$ dashed line)}
\label{A3}
\end{minipage}
\end{figure}

For the real photon the DAs of higher twists are following: the twist-3
transversal DA ($\gamma_{\mu}$ structure) (Fig. \ref{V3})
\begin{align}
\psi_{\perp\gamma}^{\left(  v\right)  }\left(  x,q^{2}=0\right)   &  =\frac
{1}{2}\left[  \delta(x)+\delta(\overline{x})-2\Theta\left(  x\overline
{x}\right)  +f_{3\gamma}^{-1}\frac{N_{c}}{4\pi^{2}}\int_{0}^{\infty}%
du\int_{-\infty}^{\infty}\frac{d\lambda}{2\pi}\frac{\left(  M_{-}%
-M_{+}\right)  ^{2}}{D_{+}D_{-}}\right]  ,\label{PsiVLinst}\\
\int_{0}^{1}dx\psi_{\perp\gamma}^{\left(  v\right)  }\left(  x,q^{2}\right)
&  =0,
\end{align}
the twist-3 DA ($\gamma_{\mu}\gamma_{5}$ structure) (Fig. \ref{A3})%
\begin{align}
\psi_{\gamma}^{\left(  a\right)  }\left(  x,q^{2}=0\right)   &  =\frac
{1}{f_{\gamma}^{\left(  a\right)  }\left(  0\right)  }\frac{N_{c}}{4\pi^{2}%
}\int_{0}^{\infty}du\int_{-\infty}^{\infty}\frac{d\lambda}{2\pi}\frac
{xM_{-}^{2}+\overline{x}M_{+}^{2}}{D_{+}D_{-}},\qquad\\
\psi_{\gamma}^{\left(  a\right)  }\left(  x=0,q^{2}=0\right)   &
=\frac{f_{3\gamma}}{f_{\gamma}^{\left(  a\right)  }\left(  0\right)  }%
,\qquad\int_{0}^{1}dx\psi_{\gamma}^{\left(  a\right)  }\left(  x,q^{2}\right)
=1,
\end{align}
the twist-4 DAs ($\sigma_{\mu\nu}$ structure)%
\begin{align}
h_{\gamma}^{\left(  t\right)  }\left(  x,q^{2}=0\right)   &  =\frac{1}%
{2}\left[  \delta(x)+\delta(\overline{x})-\frac{1}{\left\vert \left\langle
\overline{q}q\right\rangle \right\vert }\frac{N_{c}}{4\pi^{2}}\int_{0}%
^{\infty}du\int_{-\infty}^{\infty}\frac{d\lambda}{2\pi}\left(  \frac{M_{+}%
}{D_{-}}+\frac{M_{-}}{D_{+}}-\frac{\left(  M_{-}^{2}-M_{+}^{2}\right)  \left(
M_{-}-M_{+}\right)  }{D_{+}D_{-}}\right)  \right]  ,\label{hxInst}\\
\int_{0}^{1}dxh_{\gamma}^{\left(  t\right)  }\left(  x,q^{2}\right)  =0 &
,\nonumber
\end{align}%
\begin{equation}
\psi_{\gamma}^{\left(  t\right)  }\left(  x,q^{2}=0\right)  =h_{\gamma
}^{\left(  t\right)  }\left(  x,q^{2}=0\right)  .\label{gxInst}%
\end{equation}

In the case of Gaussian form factor (\ref{GaussFF} chosen in the present paper
) the integral in (\ref{hxInst}) may be partially done as%
\begin{equation}
\int_{0}^{\infty}du\int_{-\infty}^{\infty}\frac{d\lambda}{2\pi}\left(
\frac{M_{+}}{D_{-}}+\frac{M_{-}}{D_{+}}\right)  =\frac{M_{q}\Lambda^{2}}%
{2}\sum_{n=0}^{\infty}\frac{\left(  -1\right)  ^{n}}{n!}\left(  \frac
{2M_{q}^{2}}{\Lambda^{2}}\right)  ^{n}\left[  \left(  \frac{\overline{x}}%
{x}-2n\right)  ^{n}\Theta\left(  \frac{\overline{x}}{x}-2n\right)  +\left(
x\leftrightarrow\overline{x}\right)  \right]  .
\end{equation}

\begin{figure}[th]
\hspace*{0.3cm} \begin{minipage}{7cm}
\vspace*{0.5cm} \epsfxsize=6cm \epsfysize=5cm \centerline{\epsfbox
{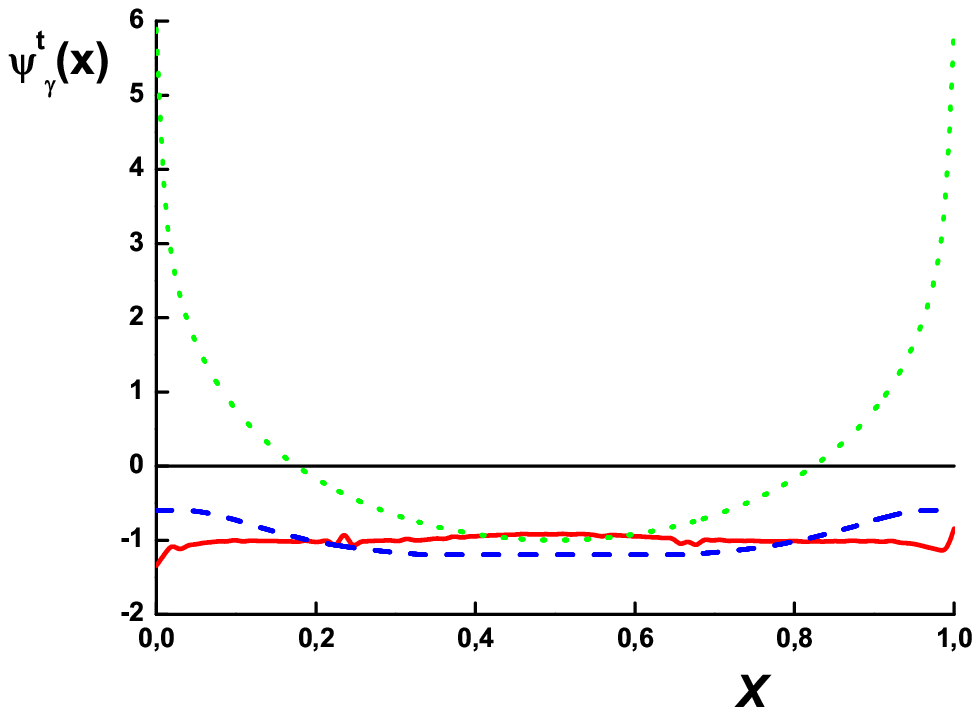}}
\caption[dummy0]{Dependence of the regular part of the twist-3 tensor component of the photon DA on transverse
momentum squared ($q^{2}=0.25$ GeV$^{2}$ solid line, $q^{2}=0$ GeV$^{2}$
dashed line, $q^{2}=-0.09$ GeV$^{2}$ short-dashed line).
The dotted line is for the WW approximation (\ref{WWt4}).           }
\label{T4}
\end{minipage}\hspace*{0.5cm} \begin{minipage}{7cm}
\vspace*{0.5cm} \epsfxsize=6cm \epsfysize=5cm \centerline{\epsfbox
{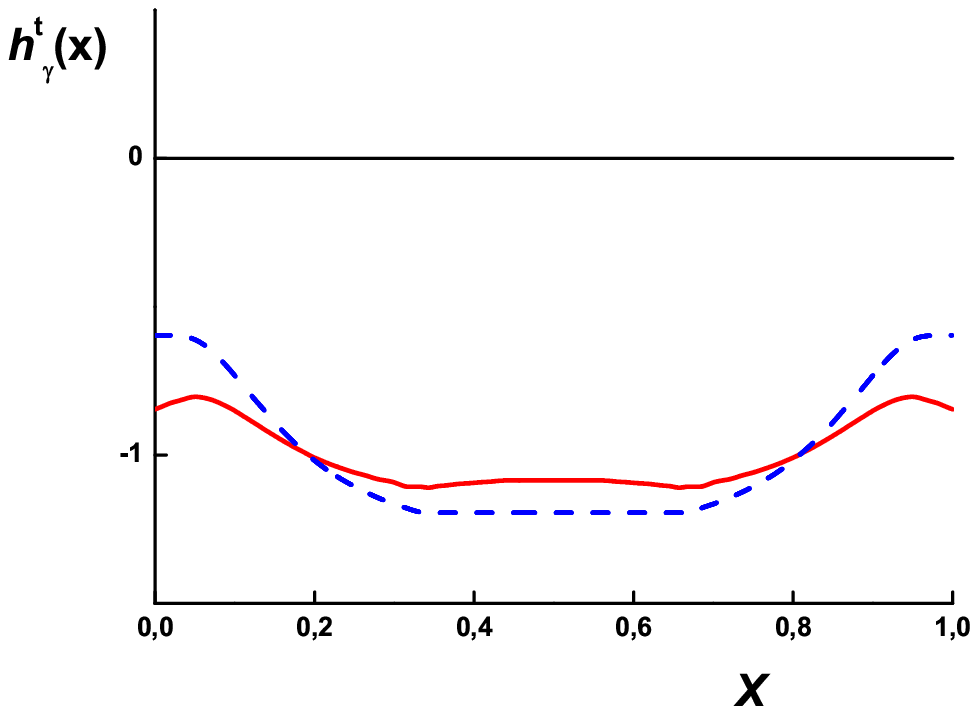}}
\caption[dummy0]{Dependence of the regular part of the twist-4 tensor component of the photon DA on
transverse momentum squared ($q^2=0.25$ GeV$^2$ solid line, $q^2=0$ GeV$^2$
dashed line)}
\label{V4}
\end{minipage}
\end{figure}

We may compare our results with the Wandzura-Wilczek (WW) approximation which
attempts to reconstruct non-leading twist result from the leading
contributions. The corresponding relations for the photon deduced from the
$\rho$-meson case \cite{Ball:1998sk} are%
\begin{align}
\psi_{\gamma}^{\left(  a\right)  \mathrm{WW}}\left(  x\right)   &  =2\left[
\int_{0}^{x}du\frac{\overline{x}\phi_{\parallel\gamma}\left(  u\right)
}{\overline{u}}+\int_{x}^{1}du\frac{x\phi_{\parallel\gamma}\left(  u\right)
}{u}\right]  ,\\
\psi_{\gamma}^{\left(  v\right)  \mathrm{WW}}\left(  x\right)   &  =\frac
{1}{2}\left[  \int_{0}^{x}du\frac{\phi_{\parallel\gamma}\left(  u\right)
}{\overline{u}}+\int_{x}^{1}du\frac{\phi_{\parallel\gamma}\left(  u\right)
}{u}\right]  ,\\
\psi_{\gamma}^{\left(  t\right)  \mathrm{WW}}\left(  x\right)   &  =\xi\left[
\int_{0}^{x}du\frac{\phi_{\perp\gamma}\left(  u\right)  }{\overline{u}}%
-\int_{x}^{1}du\frac{\phi_{\perp\gamma}\left(  u\right)  }{u}\right]  ,
\end{align}
From the fact that both the nonlocal model as well as the spectral model
predict for the leading twist photon DAs an almost constant function, we may
predict the nonleading twist DAs as they follow from WW approximation:
\begin{align}
\psi_{\gamma}^{\left(  a\right)  \mathrm{WW}}\left(  x\right)   &  =-2\left(
\overline{x}\ln\overline{x}+x\ln x\right)  ,\label{WWa}\\
\psi_{\gamma}^{\left(  v\right)  \mathrm{WW}}\left(  x\right)   &  =-\frac
{1}{2}\ln\left(  x\overline{x}\right)  ,\label{WWv}\\
\psi_{\gamma}^{\left(  t\right)  \mathrm{WW}}\left(  x\right)   &  =\xi
\ln\frac{x}{1-x}. \label{WWt4}%
\end{align}
which should be compared with the exact result obtained in the local models
\begin{equation}
\psi_{\gamma}^{\left(  t\right)  \mathrm{SQM}}\left(  x\right)  =\frac{1}%
{2}\left(  \delta\left(  x\right)  +\delta\left(  1-x\right)  \right)
-\Theta\left(  x\overline{x}\right)  .
\end{equation}
Taking literally these results one gets that the normalization conditions are
violated in the WW approximation. Indeed, all $\psi_{\gamma}^{\mathrm{WW}%
}\left(  x\right)  $ are normalized to unity. One can fix this problem by shifting the
corresponding distributions by constants, forcing the norms for $\psi_{\gamma
}^{\left(  v\right)  \mathrm{WW}}$ and $\psi_{\gamma}^{\left(  t\right)
\mathrm{WW}}\left(  x\right)$ to be zero. The corresponding distributions
are shown in Figs. \ref{V3} and \ref{T4}.

There were attempts to find photon distribution function within
next-to-leading conformal expansion with nonperturbative coefficients
determined from the QCD sum rules \cite{Ball:2002ps}. To the next-to-leading order
in the conformal expansion the twist-3 photon DA were found as%
\begin{equation}
\psi_{\gamma\perp}^{\left(  v\right)  }\left(  x\right)  =5\left(  3\xi
^{2}-1\right)  +\frac{3}{64}\left(  15\omega_{\gamma}^{V}-5\omega_{\gamma}%
^{A}\right)  \left(  3-30\xi^{2}+35\xi^{4}\right)  , \label{PsiVconf}%
\end{equation}%
\begin{equation}
\psi_{\gamma}^{\left(  a\right)  }\left(  x\right)  =\frac{5}{2}\left(
1-\xi^{2}\right)  \left(  5\xi^{2}-1\right)  \left(  1+\frac{9}{16}%
\omega_{\gamma}^{V}-\frac{3}{16}\omega_{\gamma}^{A}\right)  , \label{PsiAconf}%
\end{equation}
and the twist-4 photon DA%
\begin{equation}
h_{\gamma}\left(  x\right)  =-10\left(  1+2\varkappa^{+}\right)  C_{2}%
^{1/2}\left(  \xi\right)  , \label{Hconf}%
\end{equation}
where the parameters $\omega_{\gamma}^{V},$ $\omega_{\gamma}^{A}$ and
$\varkappa^{+}$ correspond to matrix-elements of local operators of dimension
6. In the vector-dominance approximation they are%
\begin{equation}
\omega_{\gamma}^{V}=3.8\pm1.8,\quad\omega_{\gamma}^{A}=-2.1\pm1.0,\quad
\varkappa^{+}=0. \label{ParVMD}%
\end{equation}
Note that at low scales the existing QCD sum rule results do not provide
reliable results \cite{Ball:2002ps}.

The QCD evolution of the higher-twist components of the DAs is a very
complicated problem. The reason is that unlike the leading-twist case, where
the evolution involves the single distribution and leads to simple ERBL
equations, the evolution of higher twist components couples the quark
bilinears to the three-particle quark-gluon components
\cite{Gorsky:1984js,Braun:1988qv,Ball:1998sk}. At the moment there is no
manageable procedure we could promptly use, hence we cannot repeat the
evolution calculation for the quark model prediction for non-leading twist
DAs. This important problem is interesting in its own right, but it extends
outside of the scope of this paper.

\section{Conclusions}

In the present work we have carried out an analysis of the hadronic part of
the photon DAs of leading and higher twists up to fourth order at low
normalization scale within several effective chiral quark models. We have
studied the DAs of real and virtual photons for both transverse and
longitudinal polarizations. Our analysis has been performed within two classes
of models: the nonlocal model, based on the instanton picture of the QCD vacuum, and
local models (the spectral quark model and the Nambu--Jona-Lasinio model).
Despite the different nature and details of
the calculations, our
analysis has shown a remarkable similarity of the results in both classes of models.
By working with the
photon vertex satisfying the Ward-Takahashi identity we have revised and improved the
results for the leading-twist photon DA first performed in
\cite{Petrov:1998kg}. We have shown that the calculations based on the
nonlocal model with a correct implementation of gauge invariance resulted in
much flatter results, with the photon DA remaining non-zero at the end-points
$x=0,1$, in qualitative agreement with local quark model calculations.

The main findings may be summarized as follows.

(1) The leading-twist DAs of the real photon at low momenta scale are constant in the local models
and almost constant for the nonlocal model. The independence of the photon DAs
on $x$ is expected for the local models. They also predict a constant
behaviour for the leading-twist pion DA \cite{RuizArriola:2002bp}. At the same
time the nonlocal model predicts nontrivial $x$ dependence in the case of the
pion \cite{Esaibegian:1989uj,Anikin:1999cx,Dorokhov:2002iu}, and shows a
constant behaviour for the photon independently on the shape of the nonlocal
form factor. This is a consequence of the fact that the photon has no hadronic
form factor in contrary to hadrons which are a genuine bound state of
dynamical quarks.

(2) Furthermore we show that applying the QCD evolution to leading order of
the perturbation theory, the leading-twist photon DAs become immediately zero
at the edge points of the $x$ interval (as in the pion case of
Ref.~\cite{RuizArriola:2002bp}) and are always wider (flatter) than the
asymptotic DA. At the same time the evolution of the moments of DAs is
evidently continous. This is a typical situation for functional series which
are non-uniformly convergent. The role of the QCD evolution is very important,
as the quark-model results are far from the asymptotic form. The quark model
scale estimated in Ref.~\cite{RuizArriola:2002bp} for the local models is very
low, about $320$~MeV, therefore the evolution is very fast. For the instanton
model it is estimated as $530$ MeV \cite{Dorokhov:2005pg,Dorokhov:2005hw}. In
short, our predictions may be compared to data only after a suitable QCD
evolution has been carried out. Further, for more realistic predictions the
perturbative evolution with a kernel calculated at the two-loop level
is certainly desirable.

(3) Some of the higher-twist photon DAs have delta-function behaviour at the
edge points of the $x$ interval. It might be interpreted as originating from
differentiating the constant leading twist DAs and corresponds to the
quark-anti-quark configurations of the photon DA when all momentum is carried
by one quark (or anti-quark). This singular initial behaviour is washed out
during perturbative QCD\ evolution to higher momentum scales. We also predict
the behaviour of the leading- and higher-twist DAs for nonzero virtuality of
the photon.

(4) We have drawn our attention to the fact that the equations of motion of
perturbative QCD and of the effective approaches are different. In the latter
case spontaneous breaking of chiral symmetry is taken into account in terms of
a momentum dependent dynamical quark mass. So our results do not coincide with
those obtained in perturbative QCD\ using the equations of motion. In
particular, the relations between normalization constants, and DAs for
different twist (the Wandzura-Wilczek relations) are not satisfied in the
effective approach. We question the validity of such a method, since it
implicitly assumes that current quarks are on the mass shell and free, while our
interest is in low energy matrix elements. We show that the norm of the tensor
amplitude is related to the magnetic susceptibility of the quark condensate,
and the zero norm in the vector channel is constrained by the vector-current
conservation, while the axial-vector norm remains unconstrained. The latter is
determined by nonperturbative dynamics, and therefore its value is model-dependent.

(5) Finally we compare our results with those obtained from the conformal
twist expansion \cite{Ball:2002ps} using QCD sum rules and
vector-meson-dominance methods. It seems that the results obtained in the
leading and next-to-leading orders of the conformal expansion do not converge.
We have also considered the Wandzura-Wilczek method of estimation of the
higher twist DAs assuming a constant distribution in the leading order.

\begin{acknowledgments}
AD thanks for partial support from the Russian Foundation for Basic Research
projects No. 04-02-16445, Scient. School grant 4476.2006.2 and the
JINR\ Bogoliubov--Infeld program. This research is supported by the Polish
Ministry of Education and Science, grants 2P03B~02828 and 2P03B~05925, by the
Spanish Ministerio de Asuntos Exteriores and the Polish Ministry of Education
and Science, project 4990/R04/05, by the Spanish DGI and FEDER funds with
grant no. FIS2005-00810, Junta de Andaluc\'{\i}a grant No. FQM-225, and EU RTN
Contract CT2002-0311 (EURIDICE).
\end{acknowledgments}

\appendix

\section{Useful integrals \label{app:form}}

The calculation in the nonlocal models is most involved, hence we show it
first. Formulas for the local models may then be obtained from the nonlocal
model expressions as limiting cases, with SQM involving an extra spectral integration.
We use the propagator (\ref{prop}) and
the vertex (\ref{vert1}), and evaluate the Dirac traces in (\ref{phlcwf}) for
the tensor component. The electromagnetic vertex is split into the local part
and non-local part, according to Eq.~(\ref{vert1}). We use the identity
\[
2k\cdot q=D_{+}-D_{-}+M_{+}^{2}-M_{-}^{2}.
\]
For the local part we find
\begin{equation}
\int d\tilde{k}\delta(k\cdot n-\frac{1}{2}\xi)\mathrm{Tr}[\sigma_{\alpha\beta
}S_{+}\gamma\cdot e^{(\lambda)}S_{-}]=[q_{\beta}e_{\alpha}^{(\lambda
)}-q_{\alpha}e_{\beta}^{(\lambda)}]I_{q}+[n_{\beta}e_{\alpha}^{(\lambda
)}-n_{\alpha}e_{\beta}^{(\lambda)}]I_{n},
\end{equation}
with
\begin{align}
&  \hspace{-3mm}I_{q}=4i\int d\tilde{k}\delta(k\cdot n-\frac{1}{2}\xi
)\frac{M_{+}\bar{x}+M_{-}x}{D_{+}D_{-}},\label{IqIk}\\
&  \hspace{-3mm}I_{n}=2i\int d\tilde{k}\delta(k\cdot n-\frac{1}{2}\xi)\left[
(M_{+}-M_{-})\left(  \frac{1}{D_{+}}-\frac{1}{D_{-}}\right)  \right.
\nonumber\\
&  \left.  -\frac{(M_{+}-M_{-})(M_{+}^{2}-M_{-}^{2}+q^{2}(\bar{x}-x))}%
{D_{+}D_{-}}\right]. \nonumber
\end{align}

The piece involving the nonlocal of the quark-photon interaction vertex has
the form
\begin{equation}
\int d\tilde{k}\delta(k\cdot n-\frac{1}{2}\xi)\mathrm{Tr}[\sigma_{\alpha\beta
}S(+)S(-)]M_{+,-}^{(1)}(2k-q)\cdot e^{(\lambda)}.
\end{equation}

All quantities appearing in the above integrals are functions of squares of
momenta, $k^{2}=k_{+}k_{-}-u$ and $(k-q)^{2}=(k_{+}-q_{+})(k_{-}-q_{-})-u$,
with $u=k_{\perp}^{2}$. Since the transverse and $(+-)$ spaces are factorized,
for the clarity of notation we introduce $K=(k_{0},k_{3})$ and $Q=(Q_{0}%
,Q_{3})$. In order to proceed further it is convenient to pass to the
Laplace-transform space
\begin{align}
&  F_{+}=F(k^{2})=\int_{0}^{\infty}d\alpha e^{-\alpha K^{2}}\tilde{F}_{\alpha
}(u),\label{lapl}\\
&  \hspace{-2mm}G_{-}=G((k-q)^{2})=\int_{0}^{\infty}d\beta e^{-\beta(K-Q)^{2}%
}\tilde{G}_{\alpha}(u).\nonumber
\end{align}
The terms in Eq.~(\ref{IqIk}) and in other formulas for other observables
discussed in this paper have the form of integrals over $\int\delta(k\cdot
n-\frac{1}{2}\xi)$ of functions of $k^{2}$, functions of $(k-q)^{2}$, or the
product of the two.

Using the definitions (\ref{lapl}) we find
\begin{align}
&  \hspace{-5mm}\int d^{2}K\delta(k\cdot n-\frac{1}{2}\xi)F_{+}=\int
d^{2}K\int_{-\infty}^{\infty}\frac{d\lambda}{2\pi}\int_{0}^{\infty}%
d\alpha\tilde{F}_{\alpha}(u)e^{-\alpha K^{2}-i\lambda(k\cdot n-x)}=\nonumber\\
&  \pi\int\frac{d\lambda}{2\pi}\int\frac{d\alpha}{\alpha}\tilde{F}_{\alpha
}(u)e^{i\lambda x}=\pi\int_{0}^{\infty}dz\int d\alpha e^{-\alpha z}\tilde
{F}_{\alpha}(u)\delta(x)=\pi\int dzF_{u+z}\delta(x),
\end{align}
where on the way we have carried the gaussian integration over $d^{2}K$.
Similarly,
\[
\hspace{-5mm}\int d^{2}K\delta(k\cdot n-\frac{1}{2}\xi)G_{-}=\pi\int
dzG_{u+z}\delta(\bar{x}).
\]
For integrals involving the product of $F_{+}G_{-}$ we have
\begin{align}
&  \hspace{-5mm}\int d^{2}K\delta(k\cdot n-\frac{1}{2}\xi)F_{+}G_{-}=\int
d^{2}K\int\frac{d\lambda}{2\pi}\int d\alpha\int d\beta\tilde{F}_{\alpha
}(u)\tilde{G}_{\beta}(u)e^{-\alpha K^{2}-\beta(K-Q)^{2}-i\lambda(k\cdot
n-x)}=\nonumber\\
&  \hspace{-4mm}\pi\int\frac{d\lambda}{2\pi}\int\!\!\!\frac{d\alpha d\beta
}{(\alpha+\beta)^{2}}\tilde{F}_{\alpha}(u)\tilde{G}_{\beta}(u)e^{-\frac
{\alpha\beta}{\alpha+\beta}q^{2}-i\lambda\left(  \frac{\beta}{\alpha+\beta
}-x\right)  }=\nonumber\\
&  \hspace{-4mm}\pi\int\frac{d\lambda^{\prime}}{2\pi}\int\!\!\!\frac{d\alpha
d\beta}{(\alpha+\beta)}\tilde{F}_{\alpha}(u)\tilde{G}_{\beta}(u)e^{-\frac
{\alpha\beta}{\alpha+\beta}q^{2}-i\lambda^{\prime}\left(  \bar{x}\beta
-x\alpha\right)  },\label{pos}%
\end{align}
where $\lambda^{\prime}=\lambda/(\alpha+\beta)$. Since $\int\frac
{d\lambda^{\prime}}{2\pi}\exp\left(  -i\lambda^{\prime}\left(  \bar{x}%
\beta-x\alpha\right)  \right)  =\delta(\bar{x}\beta-x\alpha)$, and $\alpha
\geq0$, $\beta\geq0$, we verify that the integral (\ref{pos}) has the correct
support$\sim\Theta(x\bar{x}).$

For the case of real photons, $q^{2}=0$, we have
\begin{align}
&  \hspace{-5mm}\int d^{2}K\delta(k\cdot n-\frac{1}{2}\xi)F_{+}G_{-}%
=\hspace{-4mm}\pi\int dz\int\frac{d\lambda}{2\pi}\int d\alpha d\beta\tilde
{F}_{\alpha}(u)\tilde{G}_{\beta}(u)e^{-(\alpha+\beta)z-i\lambda\left(  \bar
{x}{\beta}-x\alpha\right)  }=\nonumber\\
&  \hspace{-4mm}\pi\int dz\int\frac{d\lambda}{2\pi}{F}_{u+z-i\lambda x}%
{G}_{u+z+i\lambda\bar{x}}.\nonumber
\end{align}
For the general case of virtual photons one cannot unfold the Laplace
transforms. A formula convenient for numerical calculations has the form
\begin{equation}
\hspace{-5mm}\int d^{2}K\delta(k\cdot n-\frac{1}{2}\xi)F_{+}G_{-}=\pi\int
\frac{d\alpha}{\alpha}\tilde{F}_{\bar{x}\alpha}(u)\tilde{G}_{x\alpha
}(u)e^{-\alpha x\bar{x}q^{2}}.
\end{equation}

For the case on the non-local vertex we encounter integrands which are no
longer separable in the $K^{2}$ and $(K-Q)^{2}$ variables, as they $M^{(1)}$
of Eq.~(\ref{vert1}). This quantity may be written as
\[
M_{+,-}^{(1)}=\int_{0}^{1}dtM^{\prime}(tK^{2}+(1-t)(K-Q)^{2}+u),
\]
where the prime defines the derivative with respect to the argument. Hence the
Laplace transform of $M_{+,-}^{(1)}$ in the variable $\gamma$ is
$-\gamma\tilde{M}(\gamma)$.

We introduce
\[
\Delta=\alpha+\beta+\gamma,\;\;\alpha^{\prime}=\alpha+\gamma t,\;\;\beta
^{\prime}=\beta+\gamma(1-t).
\]
The following integrals are needed in our analysis:
\begin{align}
&  \hspace{-7mm}I_{1}=\int d^{2}K\delta(k\cdot n-\frac{1}{2}\xi)F_{+}%
G_{-}M_{+,-}^{(1)}k\cdot\epsilon^{(\kappa)}=\nonumber\\
&  \int\frac{d\lambda}{2\pi}\frac{d\alpha d\beta d\gamma dt}{\Delta}\tilde
{F}_{\alpha}(u)\tilde{G}_{\beta}(u)[-\gamma\tilde{M}_{\gamma}^{\prime
}(u)]e^{-\frac{\alpha^{\prime}\beta^{\prime}}{\Delta}q^{2}-i\lambda
(\beta^{\prime}\bar{x}-\alpha^{\prime}x)}\left[  -\frac{i}{2}\Delta\lambda
n\cdot\epsilon^{(\kappa)}\right] \\
&  \hspace{-7mm}I_{2}^{\mu\nu}=\int d^{2}K\delta(k\cdot n-x)F_{+}G_{-}%
M_{+,-}^{(1)}\left[  q^{\mu}k^{\nu}-q^{\nu}k^{\mu}\right]  k\cdot
\epsilon^{(\kappa)}=\nonumber\\
\int\frac{d\lambda}{2\pi}\frac{d\alpha d\beta d\gamma dt}{\Delta^{2}}  &
\tilde{F}_{\alpha}(u)\tilde{G}_{\beta}(u)[-\gamma\tilde{M}_{\gamma}^{\prime
}(u)]e^{-\frac{\alpha^{\prime}\beta^{\prime}}{\Delta}q^{2}-i\lambda
(\beta^{\prime}\bar{x}-\alpha^{\prime}x)}\times\nonumber\\
&  \left[  \frac{1}{4}(q^{\mu}\epsilon^{(\kappa),\nu}-q^{\nu}\epsilon
^{(\kappa),\mu})-\Delta^{2}\lambda^{2}(q^{\mu}n^{\nu}-q^{\nu}n^{\mu}%
)n\cdot\epsilon^{(\kappa)}\right]  .\nonumber
\end{align}
Carrying the $\lambda$ integration leads to $\delta(\beta^{\prime}\bar
{x}-\alpha^{\prime}x)$, which together with the positivity of $\alpha^{\prime
}$ and $\beta^{\prime}$ leads to the correct support (\ref{pos}).

For the case of real photons we find
\begin{align}
&  I_{1}=0,\\
&  I_{2}^{\mu\nu}=\int dz\,z\int\frac{d\lambda}{2\pi}F_{u+z-i\lambda
x}G_{u+z+i\lambda\bar{x}}M_{u+z-i\lambda x,u+z+i\lambda\bar{x}}^{(1)}\frac
{1}{4}(q^{\mu}\epsilon^{(\kappa),\nu}-q^{\nu}\epsilon^{(\kappa),\mu
}).\nonumber
\end{align}

\section{Gaussian nonlocality \label{app:Gauss}}

In this appendix we collect some useful relations valid for Gaussian shape of the nonlocal
function $f(p)$ which provides the expression for the dynamical mass as (in
chiral limit)%
\begin{equation}
M\left(  p^{2}\right)  =M_{0}\exp\left(  -2p^{2}/\Lambda^{2}\right)  .
\end{equation}
It is important to stress that exponentially decreasing function at large
Euclidean momenta is a direct consequence of requirements of the gauge invariance
and continuity of nonlocal matrix elements
\cite{Dorokhov:2000gu,Dorokhov:2005pg}. First of all consider the propagator
function%
\begin{equation}
P\left(  p^{2}\right)  =\frac{1}{p^{2}+M_{0}^{2}\exp\left(  -4p^{2}%
/\Lambda^{2}\right)  }=\frac{1}{M_{0}^{2}}\frac{1}{\widetilde{p}^{2}%
+\exp\left(  -4\widetilde{p}^{2}y^{2}\right)  },\label{propagator}%
\end{equation}
where the second equation is rewritten in the dimensionless variable $\widetilde{p}=p/M_0$, with
$y=M_{0}/\Lambda$. Depending on the values of parameters $M_{0}$ and $\Lambda$ the
function $P$ may or may not have a real pole. The critical value is obtained
from equating denominator and its derivative in $p^{2}$ to zero. Then one gets
the region of parameters where the propagator has no poles at real values of
$p^{2}$ as \cite{Bowler:1994ir,Ripka:1997zb}
\begin{equation}
y>y_{0}=1/(2\sqrt{e})\approx0.303.\label{critical}%
\end{equation}
In the other region, $y<y_0$, the real valued pole appears at%
\begin{equation}
p_{0}^{2}=-4\kappa M_{0}^{2},\label{pole}%
\end{equation}
where the coefficient $\kappa$ monotonically changes between $\kappa=1$ obtained in
the local limit $y\rightarrow0$ and $\kappa=\sqrt{e}$ reached at the critical
point  $y\rightarrow y_{0}$.

In the present paper in the nonlocal model we use the parameters
(\ref{nlparam}) corresponding to $y^{\ast}=0.195$, where the pole (\ref{pole})
appears with $\kappa=1.13$ at $p_{0}^{2}=-0.29$ GeV$^{2}$. This is a reason
why our calculations are given in the region far away from the pole. The
chosen value of the vector coupling $G_{V}$ corresponds to the $\rho$-meson mass at the
somewhat low pole
position of $M_{\rho}=541$ MeV.

\begin{figure}[tb]
\begin{center}
\includegraphics[width=7cm]{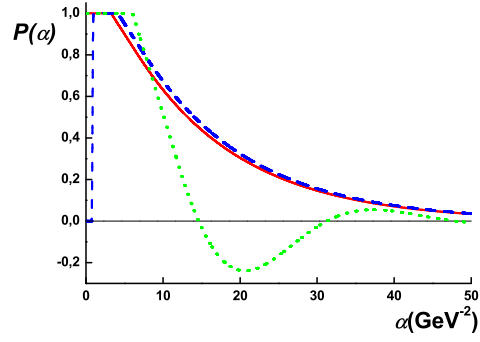}
\end{center}
\caption{The propagator functions $P$ (solid line) and $P_{f}$ (dashed line)
in $\alpha$-representation. The $\alpha$-representation of the propagator
function $P$ for "confining" set of parameters ($M_{0}=350$ MeV, $\Lambda=815$
MeV) is given by dotted line.}%
\label{alpha}%
\end{figure}

The propagator function has also an infinite number of complex valued poles
(a similar situation is also encountered in the NJL model with the proper-time
regularization \cite{Broniowski:1995yq}). This
fact makes it difficult to carry out calculations of integrals directly in the
momentum space. A big advantage is to transform all integrals to the
$\alpha$-representation where some analytical calculation may be done
\cite{Dorokhov:2000gu,Anikin:2000th}. To this end we expand (\ref{propagator})
as%
\begin{equation}
P\left(  s\right)  =\frac{1}{s}\sum_{k=-0}^{\infty}\left(  \frac{-M_{0}%
^{2}\exp\left(  -4s/\Lambda^{2}\right)  }{s}\right)  ^{k}%
\end{equation}
and apply the Laplace transfors
%yielding the function $\widetilde{P}_{f}\left(  \alpha\right)  $
\[
\frac{1}{s}=>1,\qquad\frac{1}{s^{k}}e^{-As}=>\frac{\left(  \alpha-A\right)
^{k-1}}{\left(  k-1\right)!  }\Theta\left(  \alpha-A\right)
\]
to each term of the expansion. Then one gets the $\alpha$-representation of the
propagator function
\begin{equation}
P\left(  s\right)  =>\widetilde{P}\left(  \alpha\right)  =1+\sum_{k=1}%
^{\infty}\frac{\left[  -M_{0}^{2}\left(  \alpha-4k/\Lambda^{2}\right)
\right]  ^{k}}{k!}\Theta\left(  \alpha-4k/\Lambda^{2}\right)  .\label{Palpha}%
\end{equation}
This transformation may be done for all functions. For example, one has
transformation%
\begin{equation}
P_{f}\left(  s\right)  =\frac{\exp\left(  -s/\Lambda^{2}\right)  }{s+M_{0}%
^{2}\exp\left(  -4s/\Lambda^{2}\right)  }=>\widetilde{P}_{f}\left(
\alpha\right)  =\sum_{k=1}^{\infty}\frac{\left[  -M_{0}^{2}\left(
\alpha-\left(  4k-3\right)  /\Lambda^{2}\right)  \right]  ^{k-1}}{\left(
k-1\right)  !}\Theta\left(  \alpha-\left(  4k-3\right)  /\Lambda^{2}\right)
.\label{PFalpha}%
\end{equation}
The functions $\widetilde{P}\left(  \alpha\right)  $ and $\widetilde{P}%
_{f}\left(  \alpha\right)  $ are presented in Fig. \ref{alpha}. Let us note
some general properties of the transformed functions. At small $\alpha$ one has
$\widetilde{P}\left(  0\right)  =1$ because large $s$ asymptotic $P\left(
s\rightarrow\infty\right)  =1/s$. In fact for Gaussian function $f(p^{2})$
from (\ref{Palpha}) and (\ref{PFalpha}) one has $\widetilde{P}$ $=1$ in the
interval $\alpha\in\left[  0,4/\Lambda^{2}\right]  $. At the same time
$\widetilde{P}_{f}\left(  0\right)  =0$, since at large $s$ the function $P_{f}\left(
s\right)  $ decreases faster than $1/s$. More precisely one has for Gaussian function  $\widetilde
{P}_{f}\left(  \alpha\right)  =0$ at $\alpha\in\left[  0,1/\Lambda^{2}\right]
$ and $\widetilde{P}_{f}\left(  \alpha\right)  =1$ at $\alpha\in\left[
1/\Lambda^{2},5/\Lambda^{2}\right]  $. 

The asymptotic behavior at large $\alpha$ depends on the position of the
pole closest to the real axis of the function in the $s$ space.  In the region of
parameters (\ref{critical}) it is defined by the complex pole and has
an oscillating character modulated by the exponential decay. The similar relation between
"confining" (better to say screening) properties of propagator and complex singularities of the
propagators in momentum space has been obtained in  \cite{Maris:1995ns} for QED in
three-dimensions for the fermion propagator and in \cite{Dorokhov:1999ig} for QCD
in the instanton model for the gluon field strength correlator.

For the set of parameters $y<y_{0}$ the large $\alpha$ asymptotic of
$\widetilde{P}\left(  \alpha\right)  $ is purely exponential%
\begin{equation}
\widetilde{P}\left(  \alpha\right)  \sim\exp(-\kappa M_{0}^{2}\alpha).
\end{equation}
The function $\widetilde{P}_{f}\left(  \alpha\right)  $ has the same
asymptotics because in the $s$ space it has the same dominating singularity. Through the
use of the $\alpha$-representation we avoid complexities typical for the momentum
integral calculations.

\section{QCD evolution \label{app:QCD}}

Here we collect the formulas needed for the leading-order QCD evolution. The
Gegenbauer coefficients evolve as
\begin{align}
a_{n}^{i}(Q^{2}) &  =a_{n}^{i}(Q_{0}^{2})\left(  \frac{\alpha(Q^{2})}%
{\alpha(Q_{0}^{2})}\right)  ^{(\gamma_{n}^{i}-\gamma_{0}^{i})/(2\beta_{0}%
)}\label{rat}\\
a_{n}^{i}(Q_{0}^{2}) &  =c_{n}^{m}\int_{0}^{1}dxC_{n}^{m}(2x-1)\phi
^{i}(x,Q_{0}^{2}),\label{ev2}%
\end{align}
with
\begin{align}
c_{n}^{3/2} &  =\frac{2}{3}\frac{2n+3}{(n+1)(n+2)}\nonumber\\
c_{n}^{1/2} &  =2n+1.\label{cn}%
\end{align}
The anomalous dimensions are as follows (see for example in
\cite{Shifman:1980dk}):
\begin{align}
\gamma_{n}^{\mathrm{T}} &  =-\frac{8}{3}\left[  3-4\sum_{k=1}^{n+1}\frac{1}%
{k}\right]  ,\nonumber\\
\gamma_{n}^{\mathrm{V}} &  =\gamma_{n}^{\mathrm{AV}}=-\frac{8}{3}\left[
3+\frac{2}{(n+1)(n+2)}-4\sum_{k=1}^{n+1}\frac{1}{k}\right]  ,\nonumber\\
\gamma_{n}^{\mathrm{PS}} &  =-\frac{8}{3}\left[  3+\frac{8}{(n+1)(n+2)}%
-4\sum_{k=1}^{n+1}\frac{1}{k}\right]  .\label{gambe}%
\end{align}

\end{document}